\documentclass[amsmath,amssymb,aps,twocolumn,pre,floatfix]{revtex4-2}

\usepackage{mathrsfs}
\usepackage{amsmath}
\usepackage{amssymb}
\usepackage{color}
\usepackage{graphicx}	
\usepackage{bm}
\usepackage{titlesec}
\titleformat{\paragraph}[runin]
{\bfseries\scshape}{\theparagraph}{1em}{}
\usepackage{braket}
\usepackage{multirow}
\usepackage{xcolor}

\newcommand{\be}{\begin{equation}}
\newcommand{\ee}{\end{equation}}
\newcommand{\bef}{\begin{figure}}
\newcommand{\eef}{\end{figure}}
\newcommand{\bea}{\begin{eqnarray}}
\newcommand{\eea}{\end{eqnarray}}

\begin{document}
\title{Assessing Position-Dependent Diffusion from Biased Simulations 
and Markov State Model Analysis}
\author{Fran\c cois Sicard$^{1,2}$}
\thanks{Corresponding author: \texttt{francois.sicard@free.fr}.}
\author{Vladimir Koskin$^{1,2}$}
\author{Alessia Annibale$^{3}$}
\author{Edina Rosta$^{1,2}$}
\thanks{Corresponding author: \texttt{edina.rosta@kcl.ac.uk}.}
\affiliation{$^1$ Department of Chemistry, King's College London, SE1 1DB London, UK}
\affiliation{$^2$ Department of Physics and Astronomy, University College London, WC1E 6BT London, UK}
\affiliation{$^3$ Department of Mathematics, King's College London, SE11 6NJ London, UK}
%
%
\begin{abstract}
A variety of enhanced statistical and numerical methods are now routinely 
used to extract comprehensible and relevant thermodynamic information from 
the vast amount of complex, high-dimensional data obtained from intensive 
molecular simulations. The characterization of kinetic properties, 
such as diffusion coefficients, of molecular systems with significantly 
high energy barriers, on the other hand, has received less attention. 
Among others, Markov state models, in which the long-time statistical 
dynamics of a system is approximated by a Markov chain on a discrete partition 
of configuration space, have seen widespread use in recent years, with the aim 
of tackling these fundamental issues.
Here we propose a general, automatic method to assess multidimensional 
position-dependent diffusion coefficients within the framework of Markovian 
stochastic processes and Kramers-Moyal expansion. We apply the formalism 
to one- and two-dimensional analytic potentials and data from explicit solvent 
molecular dynamics simulations, including the water-mediated conformations 
of alanine dipeptide. 
Importantly, the developed algortihm presents significant improvement 
compared to standard methods when the transport of solute across 
three-dimensional heterogeneous porous media is studied, for example, 
the prediction of membrane permeation of drug molecules.
\end{abstract}


\maketitle

\section{Introduction}
For more than a century mathematical methods have been developed to
describe the stochastic evolution of complex dynamical systems, with 
current applications in physics, chemistry, biology, 
engineering and finance~\cite{1996-Nature-Ghashghaie-Dodge,1997-PRE-Naert-Peinke,1999-PRL-Luck-Friedrich,2005-NJP-Hummer,2006-JCP-Ma-Dinner,2009-NJP-Bahraminasab-Friedrich,2010-PNAS-Best-Hummer,2013-JCTC-Comer-Gonzalez,2016-JCIM-Lee-Gumbart,2016-JCTC-Gaalswyk-Rowley}.
Of particular interest in toxicology and pharmacology is the prediction 
of kinetic quantities such as transition rates, permeation coefficients, 
and mean first-passage times of solutes, including drugs and small 
peptides, which provide fundamental understanding of numerous 
biochemical transport processes~\cite{2016-JCIM-Lee-Gumbart,2016-JCTC-Gaalswyk-Rowley,2018-JPCB-Badaoui-Rosta}.
As originally described by Hendrik A. Kramers in his seminal work~\cite{1940-Physica-Kramers}, such quantities can be calculated by
employing models of diffusive motion. In the simplest diffusive model, 
one can ignore inertia and memory effects and regard the diffusive motion 
as the random walk of a particle under a position-dependent potential~\cite{1990-RMP-Hanggi-Borkovec, 2018-PRE-Sicard}.
The dynamics of the reaction coordinate, $X(t)$, is 
determined by two functions: the potential of mean force, $W(X)$, 
and the diffusivity, $D(X)$, along this coordinate.
In general, $W(X)$ and $D(X)$ are both position-dependent 
and are likely to vary substantially in heterogeneous systems.\\

A variety of numerical methods for calculating $W(X)$ based on enhanced
sampling techniques are now well-established within the computational
community. They employ biaising potentials deposited along a few selected
degrees of freedom, named reaction coordinates (RC), to overcome the limits
of Boltzmann sampling in the presence of rare events in which energy
barriers and transition states may be poorly sampled or not sampled at all~\cite{2013-JCP-Sicard-Senet,2015-JCP-Sicard-Manghi,2016-Springer-Tiwary-vandeWalle,2018-Langmuir-Sicard-Striolo,2018-PRE-Sicard,2019-ACSNano-Sicard-Striolo,2018-APX-Camilloni-Pietrucci,2020-PRE-Sicard-Manghi}. 
The calculation of $D(X)$, on the other hand, has received less attention. 
The Einstein-Smoluchowski relation, which relates the diffusivity to the
mean square deviation of the position of the solute in the one-dimensional
long-time limit, can be used to calculate the diffusion coefficient of a
solute in a homogeneous solution by analysis of a molecular dynamics (MD)
trajectory~\cite{1956-Book-Einstein,2005-NJP-Hummer}. This relationship,
though, offers a very poor approximation of the true diffusivity in
inhomogeneous systems such as a bilayer~\cite{2016-JCTC-Gaalswyk-Rowley}. 
In these systems, the variation of the solute diffusivity is large because
the frictional environment varies dramatically as the solute moves from 
bulk water through interface, and into the membrane interior, potentially
encountering free energy barriers with heights greater than $k_BT$. 
For similar reasons, estimates based on a Green-Kubo relation of the
velocity are expected to be equally poor~\cite{1966-RPP-Kubo,2006-BC-Mamonov-Coalson}.

To circumvent this limitation, Marrink and Berendsen calculated the
diffusivity profile for the permeation of water using the force
autocorrelation function~\cite{1994-JPC-Marrink-Berendsen,1996-JPC-Marrink-Berendsen}. 
This method requires the solute to be constrained to a point
along the RC through the modification of the equation of motion of the 
MD integration, which makes it relatively difficult to apply. As a result,
it is far more convenient to perform simulations where the solute is 
simply restrained to remain near a given position with a biasing potential.
Among the latter, two strategies have been commonly employed for 
calculating $D(X)$ using biased MD simulations. The first is based on 
the generalized Langevin equation for a harmonic oscillator~\cite{2016-JCTC-Gaalswyk-Rowley,2016-JCIM-Lee-Gumbart}. 
The second employs Bayesian inferences on the likelihood of the observed
dynamics of the solute~\cite{2013-JCTC-Comer-Gonzalez,2016-JCIM-Lee-Gumbart}. 

The generalized Langevin equation provides straightforward methods to
calculate position-dependent diffusion coefficients from restrained MD
simulations. The solute is restrained by a harmonic potential so that it
oscillates at a point along the RC. The solute is then described as a
harmonic oscillator undergoing Langevin dynamics where the remainder of 
the system serves as the frictional bath for the solute. Implicitly,
describing the system as a harmonic oscillator requires the restraining
potential to be dominant over the underlying free energy surface,
\textit{i.e.} the latter is effectively a perturbation on the former. 
Values of the spring constant sufficiently large to justify this 
assumption also tend to be too large for umbrella sampling (US) simulations, 
meaning that it may not be possible to use the same simulation to calculate 
$W(X)$ and $D(X)$.
Once a time series of the $X$ position of the solute is collected, the
diffusion coefficient for that point can be calculated from the position 
or velocity autocorrelation functions (PACF and VACF, respectively). 
These methods were first introduced by Berne and co-workers~\cite{1988-JPC-Berne-Straub} 
for the calculation of reaction rates and elaborated by Woolf and Roux 
to calculate position-dependent diffusion coefficients~\cite{1994-JACS-Woolf-Roux}. 
Hummer proposed a simpler method to calculate diffusion coefficients from harmonically 
restrained simulations which avoid the need for multiple numerical 
Laplace transforms of the VACF~\cite{2005-NJP-Hummer}. However methods 
based on PACF are sensitive to the convergence of the correlation 
functions, particularly for heterogeneous bilayer environments 
where the PACF does not decay to values near zero at long time scale 
due to the lack of ergodicity in sampling~\cite{2016-JCIM-Lee-Gumbart}.
Furthermore, these algorithms are only applicable to restrained MD simulations, 
and unbiased simulations cannot be analysed with PACF or VACF.

A fundamentally different approach for the determination of
position-dependent diffusivities employs Bayesian inferences 
to reconcile thermodynamics and kinetics. In this method originally
developed by Hummer~\cite{2005-NJP-Hummer} and elaborated thereafter 
by Türkcan et al.~\cite{2012-BJ-Turkcan-Masson} and Comer 
et al.~\cite{2013-JCTC-Comer-Gonzalez}, no assumptions are made regarding 
the form of the free energy landscape on which diffusion occurs. 
This approach estimates diffusion coefficient and free energies self-consistently. 
The Bayesian scheme uses as parameters the values of the RC, $X$, along the trajectory, 
together with the force, $F(X,t)$, which is the sum of the time-dependent bias 
and the intrinsic system force, $-\nabla W(X)$. Under the stringent assumption of 
a diffusive regime, the motion is propagated using a discretized Brownian integrator. 
A Likelihood function can be constructed that gives the probability 
of observing exactly the motions along $X$ seen in the simulation runs.
Using Bayes' formula and Metropolis Monte Carlo simulations, 
the likelihood of observations given the parametrized diffusive model 
is turned into a posterior density of the unknown parameters $F(X,t)$ 
and $D(X)$ of the diffusive model, given the simulation observations. 
Additional prior knowledge and free parameters can be added to improve 
the result accuracy. However, they can make it difficult for the Bayesian
scheme to find unique solutions, and greater sampling may be needed than 
for simpler models~\cite{2005-NJP-Hummer,2013-JCTC-Comer-Gonzalez}. 
Additionally, a crucial component of this scheme is the Brownian 
integrator time step, which must be chosen to be much larger than any
correlation time of the motion~\cite{2016-JCIM-Lee-Gumbart}.\\ 

In the present study, we propose a general, automatic method for estimating
multi-dimensional position-dependent drift and diffusion coefficients
equally valid in biased or unbiased MD simulations. We combine 
the dynamic histogram analysis method (DHAM)~\cite{2015-JCTC-Rosta-Hummer}, 
which uses a global Markov state model (MSM) based on discretized RCs 
and a transition probability matrix, with a Kramers-Moyal (KM) 
expansion to calculate position-dependent drift and diffusion coefficients 
of stochastic processes, from the time series they generate, using their 
probabilistic description~\cite{1966-book-Risken,2009-PRE-Anteneodo-Riera,2009-PRE-Petelczyc-Baranowski,2009-NJP-Bahraminasab-Friedrich,2016-JORS-Rinn-Peinke,2019-JOSS-Rydin-Meirinhos}.
DHAM has been successfully employed to compute stationary quantities 
and long-time kinetics of molecular systems with significantly high 
energy barriers, for which transition states can be poorly sampled or 
not sampled at all with unbiased MD simulations~\cite{2018-JPCB-Badaoui-Rosta}. 
In this context, MSMs are extremely popular because they can be used 
to compute stationary quantities and long-time kinetics from ensembles 
of short simulations~\cite{2015-JCTC-Rosta-Hummer,2017-JCTC-Stelzl-Hummer,2017-PNAS-Olsson-Noe,2018-JACS-Husic-Pande,2018-JPCB-Biswas-Stock}.

Here we use MSMs to model MD trajectories generated from biased 
simulations. We apply the DHAM method to unbias their transition 
probability matrices and use the KM expansion to assess the original 
drift and diffusion coefficients.
We apply the formalism to one- and two-dimensional analytic potentials 
and data from explicit solvent MD simulations, including 
the water-mediated conformations of alanine dipeptide and the permeation of 
the Domperidone drug molecule across a lipid membrane. 
Our approach neither requires prior assumptions regarding the form of 
the free energy landscape nor additional numerical integration scheme. 
It can present significant improvement compared to standard methods 
when long time scale fluctuations yield a lack of ergodicity in sampling, 
as encountered in lipid bilayer systems studied in drug discovery.

\section{Methods}
\subsection{Definition of the Diffusion coefficient}
A wide range of dynamical systems can be described with a stochastic 
differential equation, the non-linear Langevin equation~\cite{2009-NJP-Bahraminasab-Friedrich,2009-PRE-Anteneodo-Riera,2014-PRE-Farago-Gronbech,2016-JORS-Rinn-Peinke,2016-SR-Bodrova-Metzler,2016-PRE-Regev-Farago}. 
Considering a one-dimensional stochastic trajectory $X(t)$ in time $t$, 
the time derivative of the system's 
trajectory $dX/dt$ can be expressed as the sum of two complementary contributions: 
one being purely deterministic and another one being stochastic, governed by 
a stochastic force. 
For a stationary stochastic process, the deterministic term is defined 
by a function $D^{(1)}(X)$ and the stochastic contribution is weighted 
by another function, $D^{(2)}(X)$, which do not explicitly depend on 
time, yielding the evolution of the equation of $X$,
\begin{equation}
\label{eqLgv}
\frac{dX}{dt} = D^{(1)}(X) + \sqrt{D^{(2)}(X)} ~\Gamma(t) \,,
\end{equation}
where $\Gamma(t)$ is a zero-average Gaussian white noise, \textit{i.e.} 
$\langle \Gamma(t) \rangle = 0$ and $\langle \Gamma(t) \Gamma(t') \rangle 
= 2\delta(t-t')$,  with $\delta$ the Dirac function. 
According to the Ito's prescription, this is equivalent to the 
Fokker-Planck equation~\cite{1966-book-Risken}
\begin{equation}
\frac{\partial P(X,t)}{\partial t}=\Big[-\frac{\partial }{\partial x} D^{(1)}(X) + \frac{\partial^2 }{\partial x^2} D^{(2)}(X) \Big]P(x,t)\,,
\label{eqFP}
\end{equation}
with stationary solution
\begin{equation}
P(X) \propto \frac{1}{D^{(2)}(X)} \exp \Bigg(\int_X dx \frac{D^{(1)}(x)}{D^{(2)}(x)} \Bigg)\,.
\label{solFP}
\end{equation}
Different integration schemes (other than the Ito's prescription) can 
be envisaged to integrate Langevin equations with a position-dependent 
diffusion coefficient, as in Eq.~\ref{eqLgv}, \textit{e.g.} the Stratonovich 
convention. Such schemes lead to different drift coefficients (often referred to 
as 'anomalous' drifts) in the Fokker-Planck equation, resulting in different 
stationary distributions~\cite{2007-PRE-Lau-Lubensky,2007-PRE-Bussi-Parrinello,2014-PRE-Farago-Gronbech,2014-JSP-Farago-GronbechJensen,2016-PRE-Regev-Farago}. \\

To infer the underlying dynamics of a stationary Markovian stochastic process, we can  
evaluate the drift and diffusion coefficients by retrieving the Kramers-Moyal 
coefficients from the time series $X(t)$~\cite{1997-PRL-Friedrich-Peinke,1998-PLA-Siegert-Peinke,2011-PR-Friedrich-Tabar}. 
The KM coefficients arise from the Taylor expansion of the master equation 
describing the Markov process, as
\begin{equation}
\label{eqKM}
D^{(n)}(X) = \lim_{\tau \to 0} \frac{1}{n!~\tau} c^{(n)}(X,\tau)  \,,
\end{equation}
which can be seen as the derivative with respect to $\tau$ of the 
$n$-th moment of the stationary conditional probability density function 
$\mathit{p}\big(X',t+\tau \mid X,t \big) = \mathit{p}\big(X',\tau \mid X \big)$~\cite{2011-PRE-Honisch-Friedrich}
\begin{equation}
\label{eqConditionalMomentContinuous}
c^{(n)}(X,\tau) = \int dX' [X'-X]^n ~\mathit{p}\big(X',\tau \mid X \big) \,.
\end{equation}
Mathematically the drift and diffusion coefficients are defined as 
the first two KM moments, \textit{i.e.} $n=1$ and $n=2$ 
in Eq.~\ref{eqKM}, respectively. 
Under the ergodic hypothesis, the average over the microstates 
defined in Eq.~\ref{eqConditionalMomentContinuous} can be equivalently 
replaced by the average over time of the trajectory $X(t)$, defined as
\begin{equation}
\label{eqConditionalMomentTimeAverage}
c^{(n)}(X,\tau) = \Big\langle [X(t+\tau)-X(t)]^n \Big\rangle_{X(t)=X} \,.
\end{equation}
Additionally, the deviation from the stochastic Langevin description given 
in Eq.~\ref{eqLgv}, \textit{i.e.}  the deviation of the driving noise 
$\Gamma(t)$ from a Gaussian distribution, can be tested with the Pawula 
theorem~\cite{1966-book-Risken}. To do so, one can compute the fourth-order 
coefficient in the KM expansion, 
$D^{(4)}(X) = \lim_{\tau \to 0} \frac{1}{4! \tau} c^{(4)}(X,\tau)$ 
and compare it to the diffusion coefficient, expecting 
$D^{(4)}(X) \ll \big(D^{(2)}(X)\big)^2$ at all $X$. 
In this work, we use MSMs to model stationary time series 
resulting from trajectories generated with a position-dependent drift $D^{(1)}(X)$ and diffusion coefficient $D^{(2)}(X)$. 

\subsection{Kramers-Moyal and Diffusion coefficients from MSM}
\begin{figure*}[t]
\includegraphics[width=0.9 \textwidth, angle=-0]{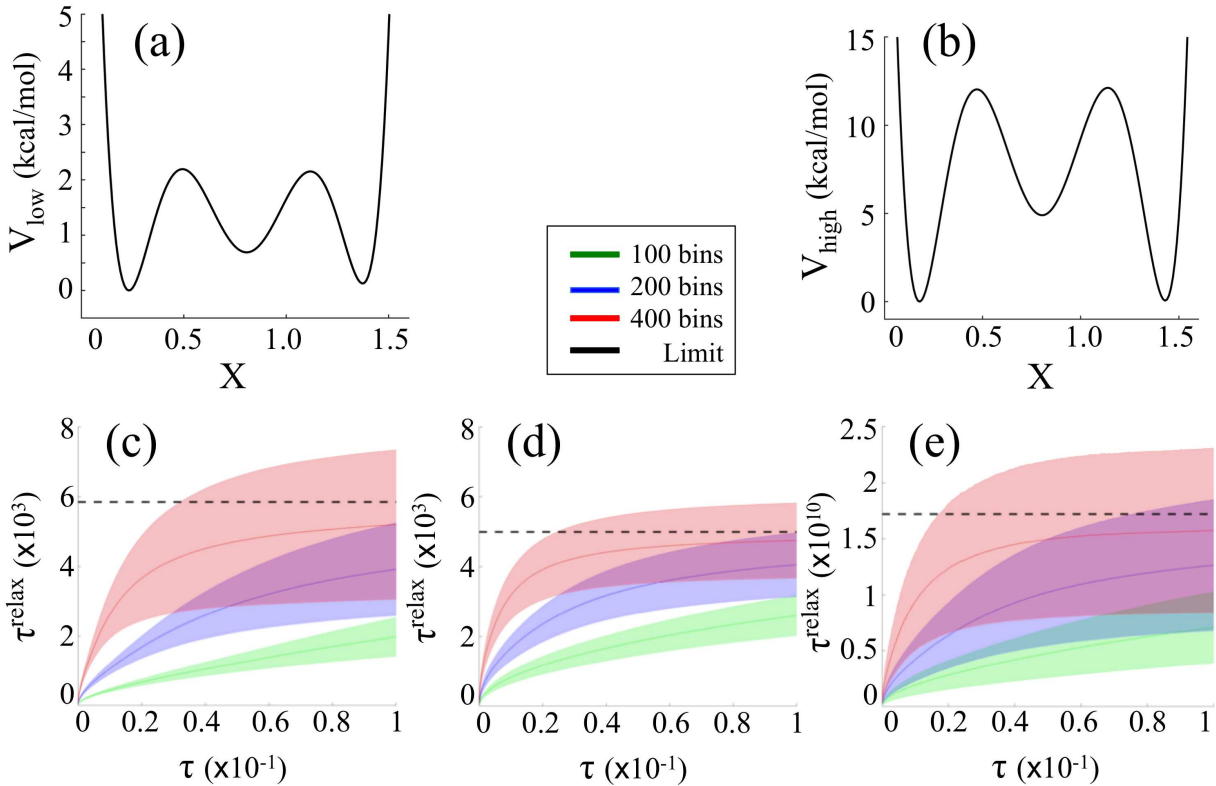}
 \caption{
 (a) Analytical potentials $V_{\textrm{low}}$ and (b) $V_{\textrm{high}}$ 
 considered in the 1D Langevin simulations.
 The lagtime dependence of the slowest relaxation time, $\tau^{\textrm{relax}}$, 
 in the overdamped Langevin dynamics is shown for (a) the unbiased simulations 
 in $V_{\textrm{low}}$, (b) the biased simulations ($K=500$ kcal/mol) 
 in $V_{\textrm{low}}$, and (c) the biased simulations ($K=500$ kcal/mol) 
 in $V_{\textrm{high}}$ for different numbers of bins, $N_{\textrm{bin}}$. 
 The limiting relaxation times, 
 $\mu^{\textrm{relax}}$, obtained  from Eq.~\ref{eqRelax} are shown with 
 dashed lines for comparison. 
 We measured $\mu^{\textrm{relax}} = 5.8\times 10^3$, $5.1\times 10^3$, 
 and $1.7\times 10^{10}$ in panels (c), (d), and (e), respectively. 
 Uncertainties, as represented in shaded area in panels (c), (d), and (e), 
 were estimated from 10 independent runs.
}
\label{fig1}
\end{figure*}

Typically, in molecular simulations we assume that the state space 
of a system evolving stochastically in time can be discretised, and this discretised 
data leads to a Markovian process. This usually involves the assumption that we have 
a low-dimensional RC, X, which describes the thermodynamical and dynamical 
evolution of the system.
In constructing MSMs, we discretise the RC, X, in $N_{\rm bin}$
bins, $\{x_i,\ldots, x_{N_{\rm bin}}\}$, defining the set of states 
(also called microstates), which the system can occupy. 
Each MD trajectory can then be analysed as a series of microstate assignments 
rather than as a series of conformations. The dynamics of the system is regarded 
as a memoryless process such that the next state of the system depends only 
on its present state. 
The number of transitions between each pair of state $i$ 
and state $j$ in the lagtime $\tau$ can then be counted 
and stored as a transition count matrix, $\mathbf{C}_{ji}^{(\tau)}$, 
which can be normalised to provide a numerical estimate of 
the transition probability matrix $M_{ji}^{(\tau)}$, at lagtime $\tau$~\cite{2010-Methods-Pande-Bowman,2011-JCP-Prinz-Schutte,2014-Springer-Prinz-Noe,2015-JCTC-Rosta-Hummer}.
Subsequently, to test the Markov condition, a spectral decomposition can 
be performed to write the Markov matrix $M_{ji}^{(\tau)}$ in terms of 
its eigenvalues and eigenvectors which provide information about the dynamics 
of the system,~\cite{2018-JCP-Kells-Rosta}
\begin{equation}
\label{eqMMspectral}
M_{ji}^{(\tau)} = \sum_n \psi_n^R(j) \psi_n^L(i) \lambda_n \,,
\end{equation}
where $\psi_n^R$ and $\psi_n^L$ are the right and left eigenvectors, respectively, 
corresponding to eigenvalues $\lambda_n$. The latter are ordered such that 
$1 = \vert\lambda_1\vert > \vert\lambda_2\vert \geq \dots \geq \vert\lambda_N\vert$. 
The second largest eigenvalue (in magnitude) of the MSMs describes the slowest relaxation 
process in the system. In practice, the slowest relaxation time, 
$\tau^{\textrm{relax}} = -\tau/\ln \lambda_2$, determined from MSMs will have 
a functional dependence on the lagtime at which the model is constructed, 
as illustrated in Figs.~\ref{fig1}(c), (d), and (e). This poses the question of 
what lagtime MSMs should be constructed at. 
Typically, we can assess the non-Markovian effects by carrying out the 
Chapman-Kolmogorov test, verifying that $n$ times the lagtime of 
a Markov process is equivalent to raising the transition matrix 
to power $n$~\cite{Bowman2014AnIT}. The relaxation time of a Markovian system 
must then be invariant under lagtime changes. The smallest lagtime $\tau$ at which 
this condition is sufficiently fulfilled is called the Markov timescale, 
$\tau_M$~\cite{2009-NJP-Bahraminasab-Friedrich,2011-PRE-Honisch-Friedrich}. \\

Given the one-dimensional time series of microstates $i\in \{1,\ldots, N_{\rm bin}\}$,
one can rewrite the $n$-th conditional moment in its discrete form
\begin{equation}
\label{eqConditionalMoment}
c^{(n)}(x_i,\tau) = \sum_{j=1}^{N_{\rm bin}} (x_j-x_i)^n ~M_{ji}^{(\tau)} \,,
\end{equation}
where $x_i$ is the (discretized) value of the RC in the center of bin $i$.
In many cases, for a given $X$, $c^{(n)}(X,\tau)$ depends linearly on $\tau$ 
in the range of $\tau$ for which the diffusive regime is satisfied. Consequently the drift 
and diffusion coefficients are estimated solely by the quotient between the corresponding 
conditional moment and the lagtime $\tau$ in this range.
Integrating Eq.~\ref{eqLgv} within the Taylor-Ito framework yields the stochastic Euler equation~\cite{2014-PRE-Farago-Gronbech,2016-PRE-Regev-Farago}
\begin{equation}
\label{eqEuler}
X(t+\tau) = X(t) + D^{(1)}(X) \tau + \sqrt{D^{(2)}(X) \tau} ~\eta(t) \,,
\end{equation}
where $\eta(t)$ is a Gaussian white noise with the same average and 
correlation as $\Gamma(t)$.
Inserting Eq.~\ref{eqEuler} into Eq.~\ref{eqConditionalMomentTimeAverage}
 yields the relations between the conditional moments and the drift and 
 diffusion coefficients~\cite{2016-JORS-Rinn-Peinke}
\begin{eqnarray}
c^{(1)}(X,\tau) & \approx & D^{(1)}(X) \tau \,, \label{eqSlope:1} \\ 
c^{(2)}(X,\tau) & \approx & 2D^{(2)}(X) \tau + \big(D^{(1)}(X) \tau\big)^2 \,. \label{eqSlope:2}
\end{eqnarray}
The latter can be subsequently assessed from the slope of a weighted 
polynomial regression of Eqs.~\ref{eqSlope:1} and \ref{eqSlope:2}.\\ 

Similarly to the one-dimensional case discussed above, the two-dimensional case comprehends 
two stochastic variables, $X(t)$ and $Y(t)$, governed by the stochastic differential equation~\cite{2009-NJP-Bahraminasab-Friedrich,2016-JORS-Rinn-Peinke}
\begin{eqnarray}
\label{eqLgv2D}
\frac{d}{dt}  \begin{bmatrix} X \\ Y \end{bmatrix} &=& 
\begin{bmatrix} D_1^{(1)}(X,Y) \\ D_2^{(1)}(X,Y) \end{bmatrix} \\
&+& \begin{bmatrix}  g_{11}(X,Y) & g_{12}(X,Y) \\ g_{21}(X,Y) & g_{22}(X,Y) \end{bmatrix} 
\begin{bmatrix} \Gamma_1(t) \\ \Gamma_2(t) \end{bmatrix}\,, \nonumber
\end{eqnarray}
where the drift coefficient $\textbf{D}^{(1)} = (D_1^{(1)}, D_2^{(1)})^T$ is 
a two-dimensional vector and the diffusion coefficient is a $2 \times 2$ matrix given 
by $\textbf{D}^{(2)} = \textbf{g}\textbf{g}^T$, with $\star ^T$ the transpose algebraic 
operator. Similar to the one-dimensional case 
the integration of Eq.~\ref{eqLgv2D} follows from a simple Euler scheme leading to 
\begin{eqnarray}
\label{eqLgv2D-2}
\begin{bmatrix} X(t+\tau) \\ Y(t+\tau) \end{bmatrix} &=& \begin{bmatrix} X(t) \\ Y(t)\end{bmatrix} +
\tau \begin{bmatrix} D_1^{(1)}(X,Y) \\ D_2^{(1)}(X,Y) \end{bmatrix}  \\ 
&+& \sqrt{\tau} \begin{bmatrix}  g_{11}(X,Y) & g_{12}(X,Y) \\ g_{21}(X,Y) & g_{22}(X,Y)\nonumber \end{bmatrix} 
\begin{bmatrix} \eta_1(t) \\ \eta_2(t) \end{bmatrix}\,.
\end{eqnarray}
The trajectory data can then be analyzed with 2D MSMs~\cite{2015-JCTC-Rosta-Hummer} 
to construct the two-dimensional Markov transition probability matrix, $\mathbf{M}^{(\tau)}$, 
at lagtime $\tau$ and the respective conditional moments yielding the expression 
for the drift and diffusion coefficients 
\begin{eqnarray}
c^{(1)}(x_i,y_i,\tau) &=&  \sum_{j=1}^{N_{\rm bin}} (x_j-x_i)~M_{ji}^{(\tau)}\,, 
\label{eqDriftDiffusion2D:1}\\
c^{(2)}(x_i,y_i,\tau) &=&  \sum_{j=1}^{N_{\rm bin}} (x_j-x_i) (y_j-y_i) ~M_{ji}^{(\tau)}\,, \label{eqDriftDiffusion2D:2}
\end{eqnarray}
which now depend on the initial bin $i$ through the values of the two 
discretised RCs, $x_i$ and $y_i$.
Extracting the drift and diffusion coefficients from MSMs and KM expansion 
depends on two parameters. 
The first parameter is the number of bins, $N_{\textrm{bin}}$, dividing 
the RC, at which $D^{(1)}$ and $D^{(2)}$ are estimated. This integer should not 
be too large that each bin does no longer include sufficient statistics
of the transition counts and also not too small that no dependence of the drift 
and diffusion on the state variable can be observed~\cite{2016-JORS-Rinn-Peinke}.
The second parameter is the range of lagtimes, $\mathbf{L}_{\tau}$, used to
build the Markov transition matrix $\mathbf{M}^{(\tau)}$
and calculate the KM coefficients 
for different $\tau$ values in Eq.~\ref{eqKM}. The conditional moments 
in Eq.~\ref{eqConditionalMoment} are computed for each bin and for each lagtime. 
For each bin, a linear fit is computed for all lagtimes in $\mathbf{L}_{\tau}$.
In practice, we have to carefully consider the range of lagtimes, 
as too short lagtimes lead to non-Markovian effects, where the approximations 
for MSMs can break down. To address the choice of lagtimes, we can also consider 
a more extensive phase space in higher dimensions, with more finely discretized MSMs, 
where non-Markovian effects are reduced.\\

The construction of MSMs from biased simulation data has not been traditionally 
possible. Biased simulations modify the potential energy function of 
the system of interest such that the system is, for example, 
harmonically restrained to a given region of the energy landscape. This is
advantageous as it allows sampling of regions which might otherwise not be adequately 
visited during the simulation time. However, the kinetic behavior observed is no longer 
representative of the true system, and as such, this needs to be accounted for 
when constructing the MSMs. 
Several recent numerical methods have been proposed in the literature 
to estimate unbiased MSMs from biased (e.g. umbrella sampling or replica exchange) 
MD simulations~\cite{2015-JCTC-Rosta-Hummer,2016-PNAS-Wu-Noe,2017-JCTC-Stelzl-Hummer,2017-JCP-Donati-Keller,2020-COSB-Kieninger-Keller}.
We used here the dynamic histogram analysis 
method (DHAM)~\cite{2015-JCTC-Rosta-Hummer}, which uses a maximum likelihood estimate of the MSM transition 
probabilities $M_{ji}^{(\tau)}$ given the observed transition counts during 
each biased trajectory. The unbiased estimate $M_{ji}^{(\tau)}$ can then be 
inserted into Eq.~\ref{eqConditionalMoment} within the KM framework, which yields 
the estimation of the drift and diffusion coefficients.

\section{Results}
In the following we give several illustrative examples of the approach 
presented in the Methods section. We used analytical models both in one and 
two dimensions within the Brownian overdamped or full inertial Langevin equations.
We also used explicit solvent simulations, including the water-mediated conformations 
of alanine dipeptide and the permeation of the Domperidone drug molecule 
across a lipid membrane. 
\begin{figure}[t]
\includegraphics[width=0.99 \columnwidth, angle=-0]{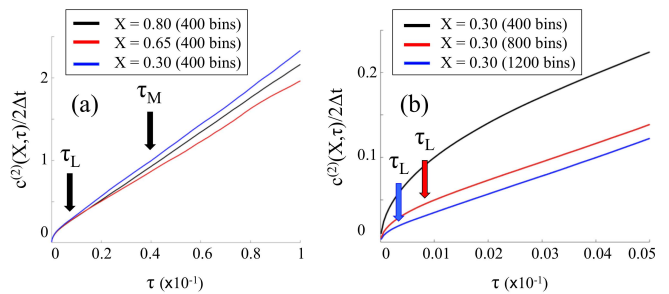}
 \caption{
 Evolution of $c^{(2)}(X,\tau)$ in the Brownian unbiased dynamics 
 and $V_{\textrm{low}}$ for (a) three different position along the RC and a 
 fixed $N_{\textrm{bin}}=400$ and (b) different values of 
 $N_{\textrm{bin}}=400$, $800$, and $1200$ and a fixed position $X=0.30$. 
 The value of the Markov timescale, $\tau_M \approx 0.04$, obtained from 
 the analysis of the relaxation time $\tau^{\textrm{relax}}$ in Fig.~\ref{fig1} 
 and $N_{\textrm{bin}}=400$ is indicated. 
 The timescale $\tau_L$ indicating the departure from the linear regime of 
 $c^{(2)}(X,\tau)$ is shown for comparaison. We measured 
 $\tau_L \approx 10^{-2}$, $10^{-3}$, and $5 \times 10^{-4}$ for 
 $N_{\textrm{bin}}=400$, $800$, and $1200$, respectively. 
 }
\label{fig2}
\end{figure}
%
\subsection{1D Brownian overdamped Langevin equation}
\begin{figure*}[t]
\includegraphics[width=0.85 \textwidth, angle=-0]{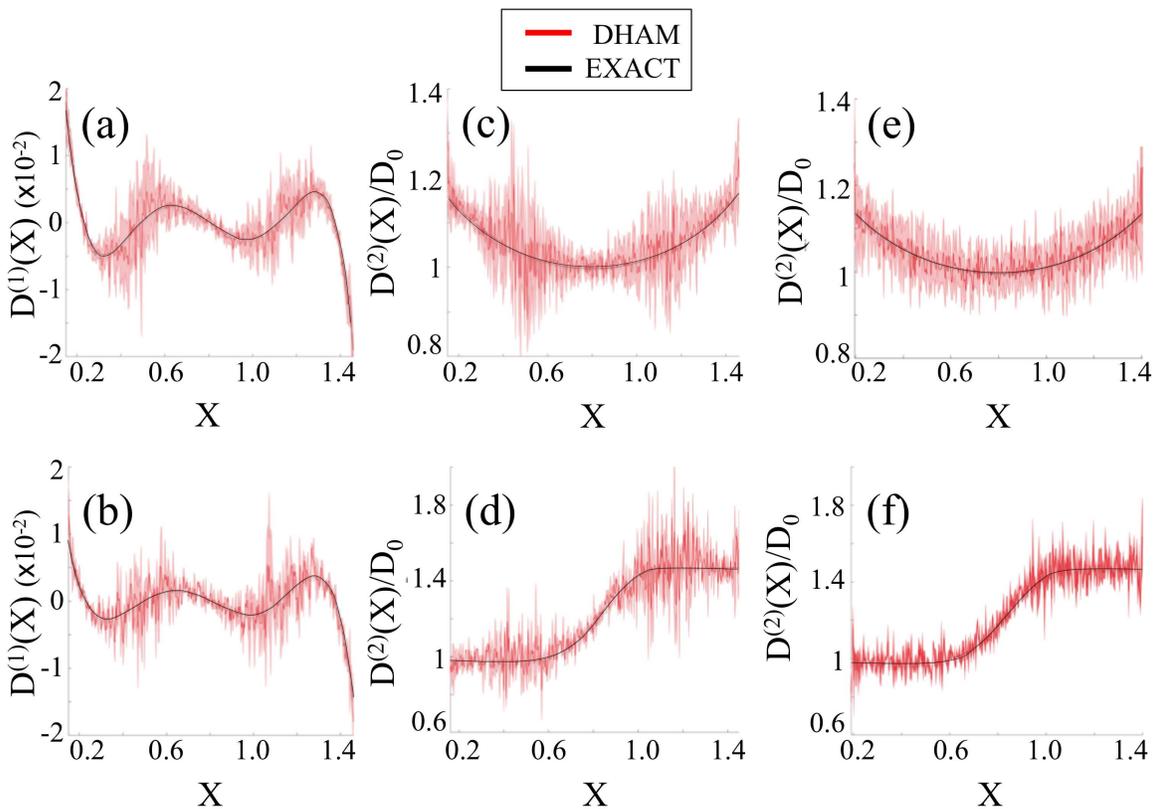}
 \caption{ 
 \textbf{Top panels.} Brownian dynamics simulations in $V_{\textrm{low}}$ 
 and quadratic diffusion profile. The drift (a) and diffusion (c) profiles 
 obtained in unbiased simulations are shown along with the diffusion 
 profile obtained in US simulations with biasing spring constant 
 $K=500$ kcal/mol (e). 
 \textbf{Bottom panels.} Brownian dynamics simulations in $V_{\textrm{low}}$ 
 and steplike diffusion profile. The drift (b) and diffusion (d) profiles 
 obtained in unbiased simulations are shown along with the diffusion 
 profile obtained in US simulations with biasing spring constant 
 $K=500$ kcal/mol (f). 
 The diffusion coefficients, $D^{(2)}(X)$, are rescaled to their respective 
 minimum values defined in the main text, $D_0 = k_BT/\gamma_0^P$ 
 or $D_0 = k_BT/\gamma_0^Z$, for clarity. 
 The profiles obtained within the DHAM framework (red) agree within standard error 
 with the exact profiles obtained from the generalization of Einstein's relation (black).
 Uncertainties, as represented in shaded area, were estimated from 10 independent runs. 
}
\label{fig3}
\end{figure*}
As a first example we integrated the 1D Brownian overdamped Langevin equation 
\begin{equation}
\frac{dx(t)}{dt} = \frac{F(x(t))}{\gamma(x(t))} + \sqrt{\frac{ k_B T}{\gamma(x(t))}}\eta(t) \,,
\label{eqBrownian}
\end{equation}
with $k_B$ the Boltzmann constant and $T=300~K$ the temperature of the system. 
In Eq.~\ref{eqBrownian}, the mass of the system is conveniently set to unity and  
$F(x)=-\nabla V(x)$ is the deterministic force derived from the 1D potential 
energy $V(x) = V_{\textrm{ref}}(x) + V_{\textrm{bias}}(x)$. 
In the following, $V_{\textrm{ref}} = \sum_{n=0}^{6} \alpha(n) x^n$ 
is defined as a polynomial of degree $6$.
We considered two different choices of the coefficients $\alpha$, 
as detailed in the supporting information (SI), leading to 
the high ($V_{\textrm{high}}$) and low ($V_{\textrm{low}}$) barrier 
potentials plotted in Fig.~\ref{fig1}(a) and (b), 
respectively.
The biased potential is defined as 
$V_{\textrm{bias}} = \frac{1}{2} K \big(x-x^{(k)}\big)^2$ 
with $K$ the biasing spring constant and $x^{(k)}$ the center of the harmonic 
bias in simulation $k$.
The parameter $\gamma(x)$ represents the position-dependent friction 
coefficient with $D(x) = k_BT/\gamma(x)$ the natural generalization 
of Einstein's relation defining the position-dependent diffusion coefficient~\cite{2014-JSP-Farago-GronbechJensen}.
We used the It\^o convention to obtain the first order 
integrator of the overdamped Langevin equation~\cite{2014-PRE-Farago-Gronbech,2016-PRE-Regev-Farago}. 
The total number of timesteps $N_{\textrm{step}}$ in the unbiased or biased 
simulation runs was chosen such that the simulation time $\Delta t \times N_{\textrm{step}} = 
5 \times 10^4$ was kept constant. 
The simulation timestep was set to $\Delta t = 10^{-5}$, which is at least an order 
of magnitude smaller than the slow characteristic time scale for the diffusion 
in the system, $\tau_{\textrm{diff}}=1/\max_x\big(\gamma(x)\big)$.\\

\textbf{Unbiased simulation.} We studied the evolution of the system 
under Eq.~\ref{eqBrownian} with the low barrier analytical potential 
$(V_{\textrm{low}})$ shown in Fig.~\ref{fig1}(a) and $V_{\textrm{bias}}=0$. 
We considered either a quadratic or step-like position-dependent diffusion 
profile with $\gamma(x)$ a parabolic (P) or a $Z$-shaped 
membership (Z) function~\cite{1998-IJAR-Medasani-Krishnapuram},
\begin{eqnarray}
    \gamma^{P}(x) &=& \gamma_0^P \Big( 1-\frac{1}{3}(x-x_P)^2 \Big) \,, \\
    \gamma^{Z}(x) &=& \gamma_0^Z \Big( 2+\textrm{zmf}(x,a,b) \Big) \,,
\end{eqnarray}
with $\gamma_0^P = 3000$, $x_P=0.8$, $\gamma_0^Z=1700$, $a=0.5$, $b=1.1$,
and $\textrm{zmf}(x,a,b)$ the sigmoidal membership function 
(see details in the SI).
We first assessed the Markov timescale $\tau_M$ associated with the trajectories 
generated via Eq.~\ref{eqBrownian} for different binning $N_{\textrm{bin}}=100$, 
$200$, and $400$, via the Chapman-Kolmogorov test. Then, we inferred the relaxation 
time of the system, in the limit of very long lagtimes, using the fitting procedure 
introduced in previous work~\cite{2018-JCP-Kells-Rosta}, which describes 
the relaxation time, $\tau^{\textrm{relax}}$, as
\begin{equation}
\label{eqRelax}
\tau^{\textrm{relax}} = \frac{\tau \times \mu^{\textrm{relax}}}{\tau + \epsilon \mu^{\textrm{relax}}} \,.
\end{equation}
In Eq.~\ref{eqRelax}, $\tau$ is the lagtime, and $\mu^{\textrm{relax}}$   
and $\epsilon$ are two free parameters which represent the true (limiting) 
relaxation timescale and the initial rate of change of the effective relaxation 
time, respectively.
As shown in Fig.~\ref{fig1}(c), increasing moderately the number of bins from $100$ 
to $400$ increases the rate of convergence towards the true relaxation time, 
$\mu^{\textrm{relax}}=5.8\times 10^3$, 
obtained from Eq.~\ref{eqRelax} with $N_{\textrm{bin}}=400$. 
We obtained similar values for the true relaxation times measured 
with $N_{\textrm{bin}}=200$ and $100$ (data not shown). 
This yields a smaller Markov timescale, $\tau_M$, which, in turn,  
gives a more accurate measure of the drift and diffusion coefficients, 
in line with the variational principle satisfied by the MSMs~\cite{2019-JCP-Kells-Rosta}.
Increasing further $N_{\textrm{bin}}$, however, each bin would eventually no longer 
include the requisite sufficient statistics for the MSM analysis. 
Considering $N_{\textrm{bin}}=400$ in Fig.~\ref{fig1}(c), the relaxation time 
can be seen to level off in the region of lagtimes greater than $0.2$. 
In the analysis that follows, we chose to define $\tau_M \approx 0.4$, 
as it is sufficiently large to be insensitive to the precise choice 
of the lagtime.
Subsequently, the first and second conditional moments were evaluated 
using Eq.~\ref{eqKM} over the range of lagtime $\tau \geq \tau_M$.

In Fig.~\ref{fig2}(a) and (b), the representative evolution 
of $c^{(2)}(X,\tau)$ is shown for different positions along the RC 
and different number of bins, respectively.
At relatively long lagtime, for $\tau \geq \tau_M$, $c^{(2)}$ 
follows the linear trend expected from Eq.~\ref{eqSlope:2}.
Close analysis of the evolution of $c^{(2)}$ shows that the linear trend 
is satisfied above a lagtime threshold, $\tau_L$, significantly lower 
than $\tau_M$.
At sufficiently short lagtime, on the other hand, for $\tau < \tau_L$, 
$c^{(2)}$ deviates from the linear trend, which stems from non-Markovian 
effects coming from the system discretization.
As shown in Fig.~\ref{fig2}(b), increasing the number of bins from 
$N_{\textrm{bin}}=400$ to $1200$ decreases the value of the lagtime threshold, 
$\tau_L$, above which the linear trend obtained in Eq.~\ref{eqSlope:2} is satisfied.

Drift and diffusion coefficients were subsequently assessed from the slope 
of a weighted polynomial regression following Eqs.~\ref{eqSlope:1} 
and \ref{eqSlope:2}. 
We also calculated the fourth-order coefficient and evaluated the ratio 
$D^{(4)}(X)/\big(D^{(2)}(X)\big)^2 < 5\times 10^{-3}$,
indicating the validity of the condition of the Pawula theorem. 
As shown in Fig.~\ref{fig3}(a)-(d), 
we observed excellent agreement between the theoretical profiles and 
the numerical results for both position-dependent drift and diffusion 
coefficients, with higher variability around $X \approx 0.5$ 
and $\approx 1.2$, where the derivative of the low barrier potential  
$V_{\textrm{low}}$, shown in Fig.~\ref{fig1}(a), is maximal.\\
%

\textbf{Biased simulation.} We extented the previous analysis to the biased evolution 
of the system in the low $(V_{\textrm{low}})$ and high $(V_{\textrm{high}})$
barrier potentials shown in Figs.~\ref{fig1}(a) and (b), respectively, 
within the US framework. 
We ran standard US simulations with the biasing potential 
$V_{\textrm{bias}}$ defined above in each umbrella window.
We used $50$ uniformly distributed umbrella windows in the range $[0.25, 1.35]$ 
along the RC. This number was sufficiently high to obtain accurate sampling, 
given the different biasing spring constants considered in this work.
\begin{figure}[t]
\includegraphics[width=0.99 \columnwidth, angle=-0]{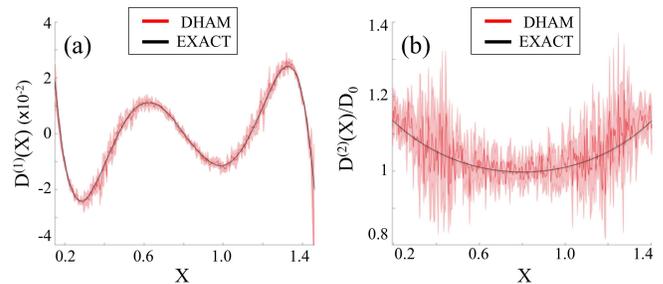}
 \caption{Drift (a) and diffusion (b) profiles obtained in biased 
 overdamped Langevin simulations for the high energy barrier potential 
 ($V_{\textrm{high}}$). A biasing spring constant $K=500$ kcal/mol and a quadratic 
 diffusion profile are used. The diffusion coefficient $D^{(2)}(X)$ is rescaled to 
 $D_0 = k_BT/\gamma_0^P$ for clarity.
 The profiles obtained within the DHAM framework (red) agree within standard error 
 with the exact profiles obtained from the generalization of Einstein's relation (black).
 Uncertainties, as represented in shaded area, were estimated from 10 independent runs. 
}
\label{fig4}
\end{figure}

We first studied the effect of the bias on the reconstruction of the diffusion 
coefficient when the system evolves in $V_{\textrm{low}}$. 
We set the biasing spring constant to $K=500$ kcal/mol, which was sufficienlty 
strong to allow good sampling in both low ($V_{\textrm{low}}$) 
and high ($V_{\textrm{high}}$) barrier potentials.
The lagtime dependence of $\tau^{\textrm{relax}}$ is shown in Fig.~\ref{fig1}(d). 
We observed a similar behaviour 
to the one obtained in the unbiased simulations (Fig.~\ref{fig1}(c)) with the decline of 
the non-Markovian effects on the dynamics when the  number of bins, 
$N_{\textrm{bin}}$ increases from $100$ to $400$. Most noticeably, 
the use of the biasing spring constant decreases significantly 
the variability of the relaxation timescale at longer lagtimes.
As shown in Figs.~\ref{fig3}(e) and (f), excellent agreement is observed between the theoretical 
profiles and the numerical results. 
We also confirmed that $D^{(4)}(X)/\big(D^{(2)}(X)\big)^2 < 5\times 10^{-3}$.
As expected, the use of the biasing spring constant yields better sampling 
across the RC.

We complemented this analysis with the reconstruction of the diffusion 
coefficient of the system using $V_{\textrm{high}}$ and a biaising 
spring constant $K=500$ kcal/mol. 
The relaxation timescale is shown in Fig.~\ref{fig1}(e). 
We observed a similar behavior to the  unbiased simulations 
with the decline of the non-Markovian effects on the dynamics when the 
number of bins, $N_{\textrm{bin}}$ increases from $100$ to $400$. 
The  relaxation time of the system calculated for $N_{\textrm{bin}}=400$ 
converges towards a limiting value, $\mu^{\textrm{relax}}=1.7\times 10^{10}$, 
significantly higher than the one measured for $V_{\textrm{low}}$, 
due to the longer equilibration time needed to cross the high energy barrier.
As shown in Figs.~\ref{fig4}(a) and (b), we observed good agreement 
between the theoretical profiles and the numerical results for 
the drift and diffusion coefficients. 
We also verified that $D^{(4)}(X)/\big(D^{(2)}(X)\big)^2 < 5\times 10^{-3}$.
We noticed, however, higher variability 
around $X \approx 0.5$ and $\approx 1.2$, where the derivative of 
the high barrier potential $V_{\textrm{high}}$ (Fig.~\ref{fig1}(b)) is maximal. 
As shown in the SI, the increase of the biasing spring constant 
from $500$ kcal/mol to $2000$ kcal/mol yields 
better sampling around the transition states with lower variability on 
the reconstruction of the diffusion profile, as already observed 
in the low barrier case.

\subsection{1D full inertial Langevin equation}
\begin{figure}[t]
\includegraphics[width=0.99 \columnwidth, angle=-0]{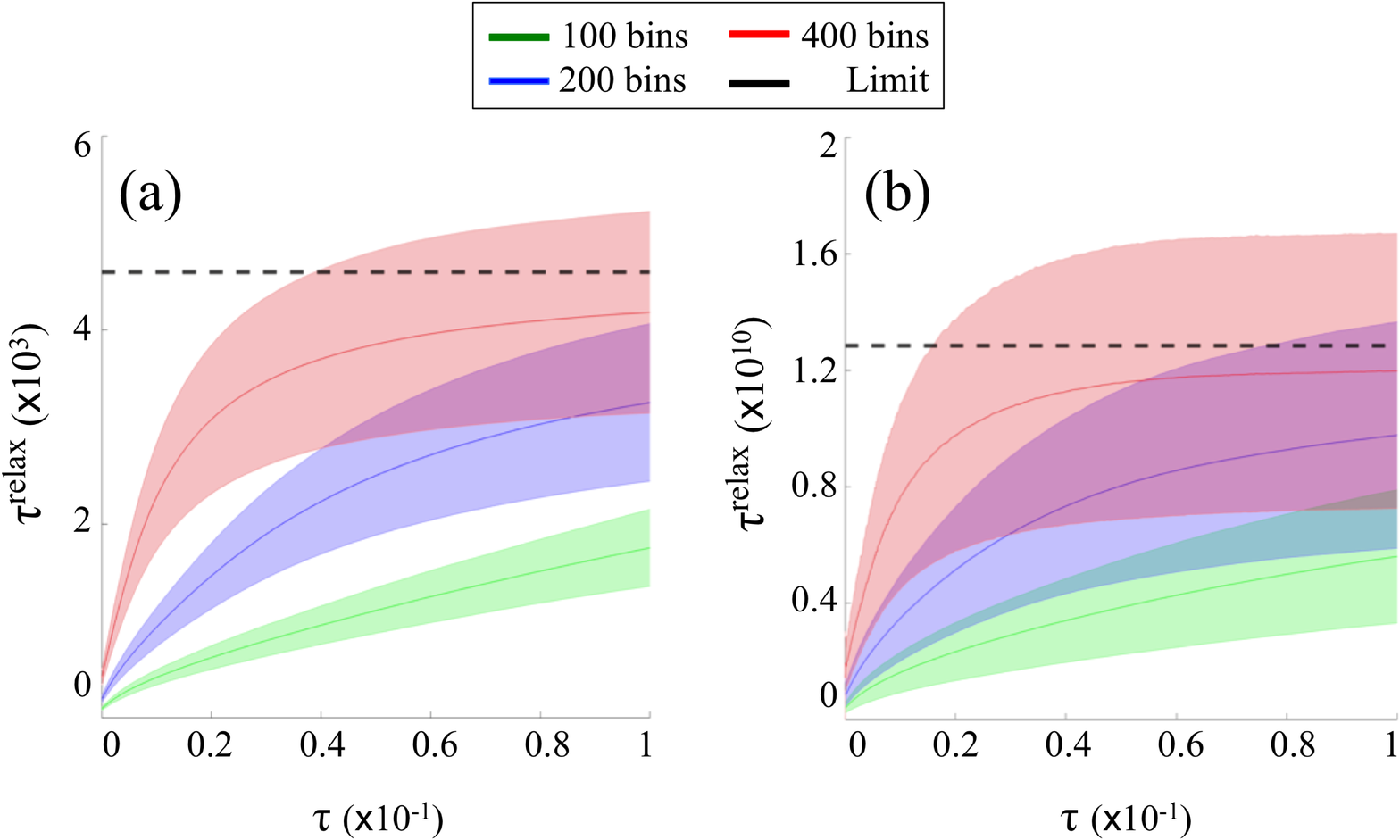}
\includegraphics[width=0.99 \columnwidth, angle=-0]{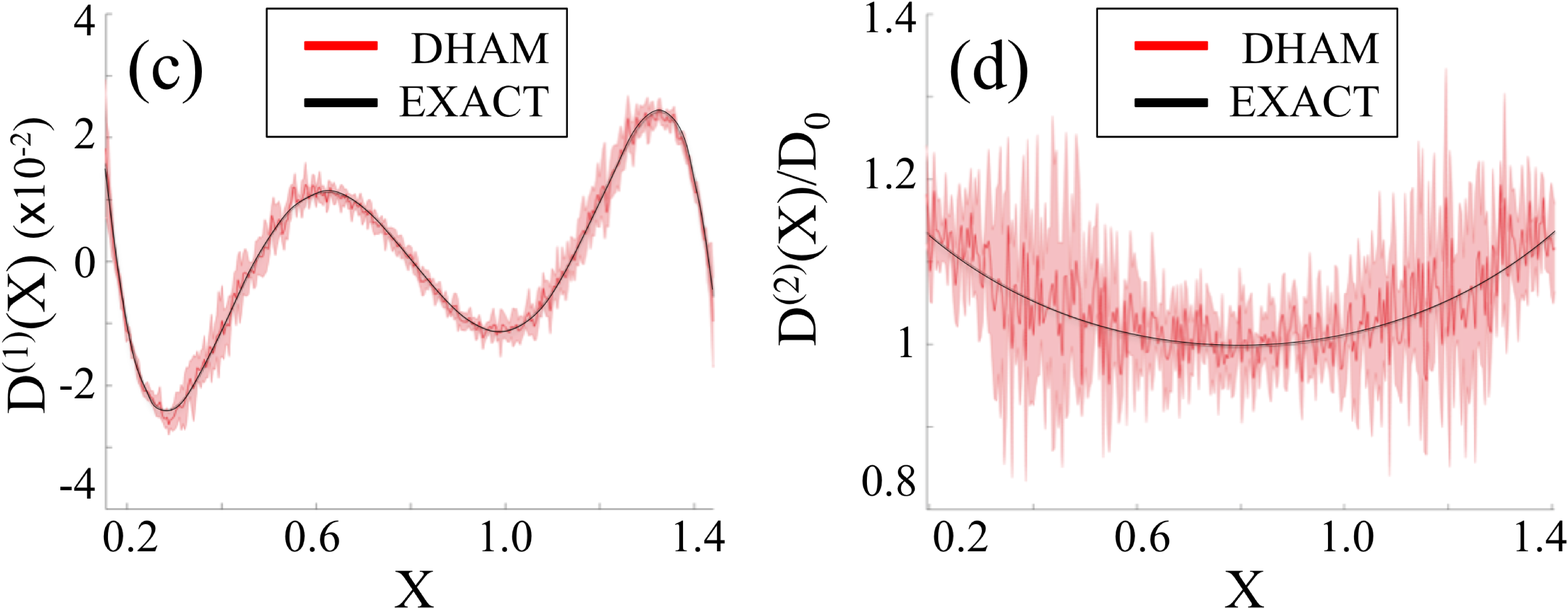}
 \caption{
Relaxation times $\tau^{\textrm{relax}}$ as a function of the lagtime, $\tau$, 
and the number of bins, $N_{\textrm{bin}}$, obtained in biased 1D full Langevin 
simulations for (a) low barrier ($V_{\textrm{low}}$) and (b) high barrier 
($V_{\textrm{high}}$) energy  potentials. A biasing spring 
constant $K=500$ kcal/mol is used. Drift (c) and diffusion (d) obtained 
for US simulations with biasing spring constant $K=500$ kcal/mol and quadratic 
diffusion profile. The diffusion coefficient $D^{(2)}(X)$ 
is rescaled to $D_0 = k_BT/\gamma_0^P$ for clarity.
The profiles obtained within the DHAM framework (red) in panels (c) and (d) 
agree within standard error with the exact profiles obtained from the generalization 
of Einstein's relation (black). 
Uncertainties, as represented in shaded area, were estimated from 10 independent 
runs.
}
\label{fig5}
\end{figure}
To take into account the role played by inertia in the reconstruction 
of the drift and diffusion coefficients, we extended Eq.~\ref{eqBrownian}  
to the full inertial Langevin equation
\begin{equation}
\frac{d^2x(t)}{dt^2} = F(x(t)) - \gamma(x(t))\frac{dx(t)}{dt} + \sqrt{k_B T \gamma(x(t))}\eta(t) \,,
\label{eqFull}
\end{equation}
where the mass of the system is conveniently set to unity. 
From this relation, we used the Vanden-Eijnden and Ciccotti algorithm~\cite{2006-CPL-VandenEijnden-Ciccotti} that generalises the Velocity Verlet 
integrator to Langevin dynamics, along with the It\^o convention. This scheme takes 
the inertial term into account and is accurate to order $\Delta t^2$.

We used the same simulation timestep, $\Delta t = 10^{-5}$,  
and biasing spring constants, $K=500$ kcal/mol,  as in the Brownian overdamped 
Langevin dynamics.
The lagtime dependence of the relaxation time $\tau^{\textrm{relax}}$, 
as shown in Fig.~\ref{fig5}(a) and (b) for $V_{\textrm{low}}$ and $V_{\textrm{high}}$, 
respectively, is similar to the one measured in 
the Brownian overdamped Langevin simulation, with the decline of the  
non-Markovian effects when the  number of bins, $N_{\textrm{bin}}$ increases 
from $100$ to $400$, and the convergence towards the limiting relaxation times 
$\mu^{\textrm{relax}}=4.4\times 10^3$ and $1.3\times 10^{10}$ for 
$V_{\textrm{low}}$ and $V_{\textrm{high}}$, respectively, calculated 
for $N_{\textrm{bin}}=400$.

We limited the rest of the analysis to the biased evolution of the system 
in $V_{\textrm{high}}$. 
The associated drift and diffusion profiles are shown in Fig.~\ref{fig5}(c) 
and (d) (see details in the SI). We observed good agreement between theoretical 
and numerical profiles, with $D^{(4)}(X)/\big(D^{(2)}(X)\big)^2 < 4\times 10^{-3}$.
As already noted in the overdamped Langevin dynamics, we observed higher variability 
around $X \approx 0.5$ and $\approx 1.2$, where the derivative of the potential energy 
$V_{\textrm{high}}$ is maximal. Increasing the biasing spring constant 
from $K=500$ kcal/mol to $2000$ kcal/mol improved the accuracy of the measure 
of the diffusion coefficient (Fig.~S2).

\subsection{2D Brownian overdamped Langevin equation}
\begin{figure}[t]
\includegraphics[width=0.99 \columnwidth, angle=-0]{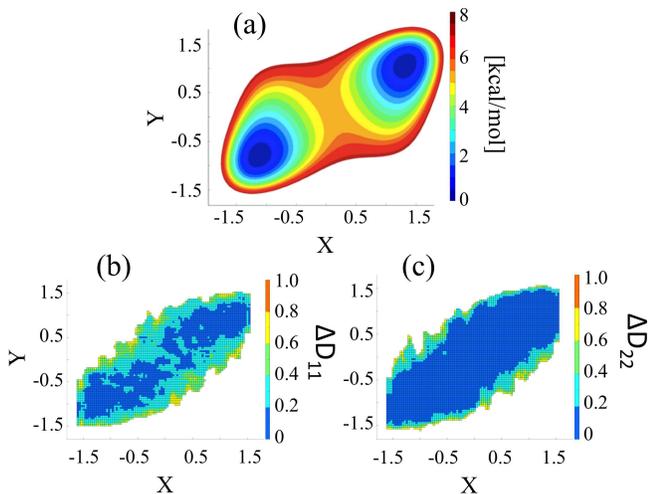}
 \caption{ (a) Free-energy surface associated with the 2D analytical 
 model potential projected along the $x$- and $y$-axes.
 The 2D projections of the deviation of the numerical diagonal elements 
 of the diffusion tensor, $\Delta D_{11}$ and $\Delta D_{22}$, are shown in 
 panels (b) and (c), respectively. 
 }
\label{fig8}
\end{figure}

We extended the 1D analysis above to the estimation 
of the 2D diffusion tensor of the analytical system evolving under 
the 2D Brownian overdamped equation~\cite{2000-book-Freund-Poschel}
\begin{equation}
\gamma(t)\frac{d\mathbf{X}}{dt}=\mathbf{F}(\mathbf{X}) + \sqrt{k_BT}~\mathbf{b}(t)~\eta(t)
\label{eq2DLgv1}
\end{equation}
where $\mathbf{X}\equiv \big(X,Y\big)$ is a 2D vector and  
$\textbf{F}(\mathbf{X}) = -\nabla V(\mathbf{X})$ is the deterministic force 
derived from the 2D potential energy $V(\mathbf{X})$. 
In Eq.~\ref{eq2DLgv1}, the force $\textbf{b}(t)$ is related to the friction 
tensor $\mathbf{\gamma}(t)$ as follows
\begin{equation}
\mathbf{\gamma}(t) = \textbf{b}(t)\textbf{b}^T(t)\,,
\end{equation}
and $\eta(t)=\big(\eta_1(t), \eta_2(t)\big)$ is a 2D vector of independent, 
identical Gaussian white noise sources, \textit{i.e.} $\langle \eta_i(t) \rangle = 0$ 
and $\langle \eta_i(t)\eta_j(t') \rangle = 2\delta_{i,j}\delta(t-t')$.
From this relation, we used the It\^o convention to obtain the first order 
integrator of Eq.~\ref{eq2DLgv1}.
For economy of computational resources, we focused our analysis 
on a diagonal friction tensor 
\begin{equation}
\gamma(t) = \begin{bmatrix}  \gamma_{11}(\mathbf{X}) & 0 \\ 0 & \gamma_{22}(\mathbf{X}) \end{bmatrix}
\end{equation}
with constant friction $\gamma_{11}(\mathbf{X}) = 300$ and 
$\gamma_{2}(\mathbf{X}) = 30$ and the 2D analytical potential 
$V(\mathbf{X}) = -3 X^2 + X^4 -3X Y + Y^4$ shown in Fig.~\ref{fig8}(a). 
We modeled standard US simulations, evenly positioned along the RC, 
with a 1D bias potential 
$V_{\textrm{bias}}(\mathbf{X}) = V^{(k)}(X) = \frac{1}{2} K (X-X^{(k)})^2$ 
in each umbrella 
window $k$, with $K$ the biasing spring constant and $X^{(k)}$ the center of 
the harmonic bias in window $k$. We used $50$ uniformly distributed umbrella windows 
in the range $[0.25, 1.35]$ along the $x$-axis. 
We then constructed a discretized 2D grid 
to determine the MSMs along the $x$- and $y$-axes. 

The diagonal elements of the diffusion tensor were obtained from the analysis 
of the KM coefficients defined 
in Eqs.~\ref{eqDriftDiffusion2D:1} and \ref{eqDriftDiffusion2D:2} 
(see details in the SI). 
We also calculated the fourth-order coefficients and verify that the ratio 
$D_{11}^{(4)}(X)/\big(D_{11}^{(2)}(X)\big)^2 < 5\times 10^{-3}$ and 
$D_{22}^{(4)}(X)/\big(D_{22}^{(2)}(X)\big)^2 < 6\times 10^{-4}$, 
indicating the validity of the condition of the Pawula theorem. 
To assess the accuracy of the reconstructed diffusion coefficient, 
we measured the deviation between the observed (ob) and  theoretical (th) 
values
\begin{equation}
    \Delta D_{ii}(\mathbf{X}) = \frac{\big\vert D_{ii}^{(\textrm{ob})}(\mathbf{X}) - D_{ii}^{(\textrm{th})}(\mathbf{X})\big\vert}{D_{ii}^{(\textrm{th})}(\mathbf{X})} \,,
\end{equation}
with $D_{ii}^{(\textrm{th})}(\mathbf{X}) = k_BT/\gamma_{ii}(\mathbf{X})$.
In Fig.~\ref{fig8}(b) and (c), it is shown the numerical measures 
for $\Delta D_{11}(\mathbf{X})$ and $\Delta D_{22}(\mathbf{X})$, respectively. 
The observations are in good agreement with the theoretical values 
$D_{11}^{\textrm{th}}(\mathbf{X}) = 2\times 10^{-3}$ and $D_{22}^{\textrm{th}}(\mathbf{X}) = 2\times 10^{-2}$. 
Most noticeably, the lower value of the friction parameter $\gamma_{22}(\mathbf{X})$ 
allows better sampling, which yields a more accurate measure for $D_{22}(\mathbf{X})$.
Finally, to quantify the accuracy of the reconstruction, we measured the effective 
diffusion coefficient along each reaction coordinate
\begin{equation}
D_{ii}^{\textrm{eff}} = \frac{1}{N_{\textrm{bin}}}\int D_{ii}(\mathbf{X})~d\mathbf{X} \,.
\label{Deff2D}
\end{equation}
We measured $D_{11}^{\textrm{eff}}=(1.9 \pm 0.4)\times 10^{-3}$ 
and $D_{22}^{\textrm{eff}}=(2.0 \pm 0.7)\times 10^{-2}$, 
in good agreement with $D_{11}^{\textrm{th}}$ and $D_{22}^{\textrm{th}}$, 
respectively.
\begin{figure}[b]
\includegraphics[width=0.99 \columnwidth, angle=-0]{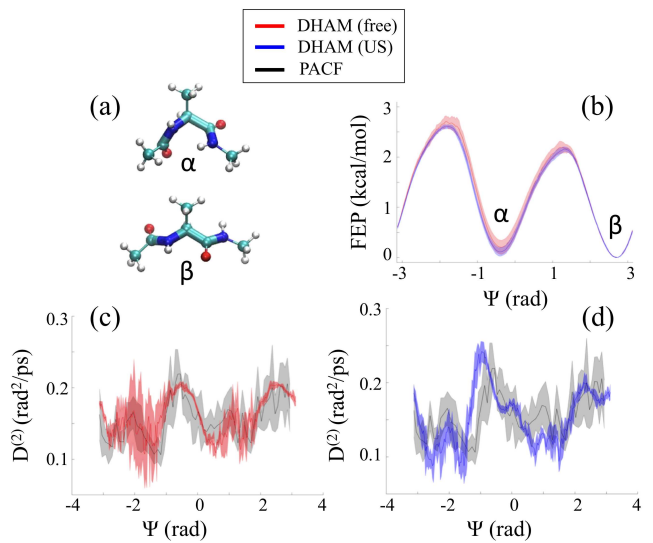}
 \caption{(a) Ball-and-stick representation of the conformers 
 $\alpha$ and $\beta$ of Ala2 along with (b) the free energy 
 profile reconstructed along the dihedral angle $\Psi$ $(N1-C\alpha-C2-N2)$ using the DHAM 
 in unbiased (red) and biased (blue) simulations. The diffusion profiles obtained in 
 (c) unbiased and (d) biased simulations agree within standard errors 
 with the one obtained within the PACF method of Hummer (grey)~\cite{2005-NJP-Hummer}. 
 Uncertainties, as represented in shaded areas, were estimated from 3 independent 
 runs.
}
\label{fig6}
\end{figure}
%

\subsection{Water-mediated conformations of Alanine Dipeptide in 1D}
The conformational transition between the different conformers of the 
solvated alanine dipeptide (Ala2) has been extensively used as a case study 
for several theoretical and computational investigations~\cite{2005-JCP-Ren-Weinan, 2010-JPCB-Vymetal-Vondrasek,2011-JCTC-Yonezawa-Nakamura,
2013-PRL-Tiwary-Parrinello,2017-JMM-Cuny-Mineva,2018-PRE-Sicard}. 
We studied the transition between the metastable states $\alpha$ and $\beta$ of 
Ala2 shown in Fig.~\ref{fig6}(a), which can be differenciated by the values of the 
backbone dihedral angle $\Psi$ and are separated by an activation free energy barrier 
of $\approx 2$ kcal/mol at the temperature $T=300~K$. 
We used a Langevin thermostat to enforce the temperature~\cite{2007-JCP-Bussi-Parrinello}, 
a time step of $0.2$ fs, AMBER03 force field~\cite{2005-JCC-Case-Woods} with TIP3P water 
model~\cite{1983-JCP-Jorgensen-Klein}, and GROMACS 5.1 molecular dynamics 
code~\cite{2001-JMM-Lindahl-VanDerSpoel}. A single alanine dipeptide molecule was kept 
in a solvated periodic cubic box of edge $\approx 3$ nm. The LINear Constraint Solver (LINCS) 
algorithm~\cite{1997-JCC-Hess-Fraaije} handled bond constraints while the particle-mesh Ewald 
scheme~\cite{1993-JCP-Darden-Pedersen} was used to treat long-range 
electrostatic interactions. The non-bonded van der Waals cutoff radius was $0.8$ nm.

The relatively low energy barriers allow the system to be sampled with both unbiased 
and biased simulations, which gives a mean to assess the accuracy of the results 
obtained with enhanced sampling methods.
We determined the free energy profile and diffusion coefficient by using either 
free simulations ($500$ ns) or US biased simulations and the DHAM approach. 
The starting structures for the US simulations were obtained 
by pulling the system along the dihedral angle $\Psi$ within the range $[-\pi,\pi]$. During the 
simulations, a snapshot was saved every $0.1$ rad generating $60$ windows. Each US window was 
subsequently run for $1$ ns to allow equilibration, followed by additional $5$ ns of the production 
run using an US force constant of $5$ kcal~$\textrm{mol}^{-1}$~$\textrm{rad}^{-2}$.
In Fig.~\ref{fig6}(b), (c), and (d) are shown the reconstructed FEPs 
and diffusion coefficients obtained within the DHAM framework, 
either with unbiased (red) or biased (blue) simulations. 
Both approaches gave similar results. 
In particular, we compared the diffusion profiles obtained with DHAM 
with the one obtained within the PACF framework (grey),
\begin{equation}
D(\Psi_k=\langle \Psi \rangle_k) = \frac{\textrm{var}(\Psi)^2}{\int_0^\infty C_{\Psi}(t)~dt}~.
\label{Diffeq}
\end{equation}
In Eq.~\ref{Diffeq}, $\langle \Psi \rangle_k$ is the average of the RC in the US window $k$, 
$\textrm{var}(\Psi)=\langle \Psi^2 \rangle - \langle \Psi \rangle^2$ is its variance, and 
$C_{\Psi}(t)=\langle \delta \Psi(0) \delta \Psi(t) \rangle$ the PACF calculated directly 
from the time series. 
Following the original work of Hummer, we increased the strength of the biasing 
potential from $\approx 5$ kcal~$\textrm{mol}^{-1}$~$\textrm{rad}^{-2}$ 
to $\approx 24$ kcal~$\textrm{mol}^{-1}$~$\textrm{rad}^{-2}$ to satisfy 
the underlying assumption that the harmonic restraint be sufficiently large to render 
the underlying free energy surface a small perturbation on the harmonic 
potential~\cite{2005-NJP-Hummer}. 
As shown in Fig.~\ref{fig6}(c) and (d), we observed good agreement between the results 
derived with DHAM either in the unbiased or biased simulations and 
the PACF method. To compare the results with those in the literature, 
we measured the effective diffusion coefficient 
\begin{equation}
D^{(2)}_{\textrm{eff}} = \frac{1}{N_{\textrm{bin}}}\int_{-\pi}^{+\pi} D^{(2)}(\Psi)~d\Psi\,.
\label{D2eff}
\end{equation}
with $N_{\textrm{bin}}$ the number of bins used in the MSM. 
Eq.~\ref{D2eff} yields the effective Diffusion coefficient 
$D^{(2)}_{\textrm{eff}} \approx 0.16~\textrm{rad}^2/\textrm{ps}$, 
in  agreement with the result obtained by Hummer et al.~\cite{2003-JCP-Hummer-Kevrekidis,2005-NJP-Hummer} 
($\approx 0.15~\textrm{rad}^2/\textrm{ps}$) and Ma et al.~\cite{2006-JCP-Ma-Dinner} 
($\approx 0.34~\textrm{rad}^2/\textrm{ps}$).

\subsection{Membrane permeation in 1D}
\begin{figure}[b]
\includegraphics[width=0.99 \columnwidth, angle=-0]{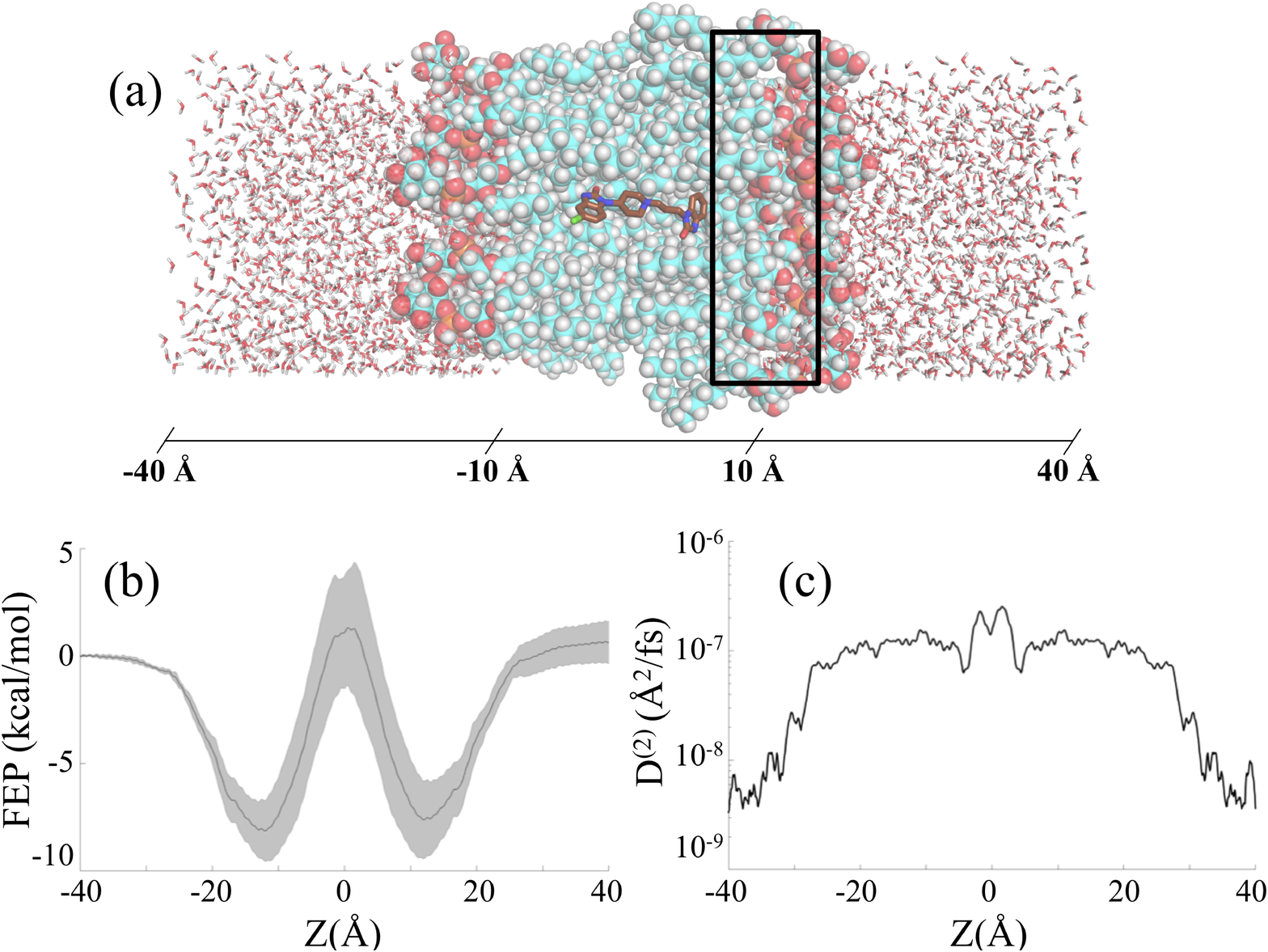}
 \caption{
 (a) Schematic representation of Domperidone (in brown at the center 
 of the image) at the center of a POPC bilayer surrounded by water. The local area 
 centered around $Z=10 \AA$, considered in the 2D calculation of the diffusion tensor, 
 is shown (solid black rectangle).
 (b) Free energy and (c) diffusion profiles (in semi-log representation) associated 
 with the crossing of the drug molecule through the bilayer and reconstructed 
 along the distance from the center of the bilayer, $Z$. 
Uncertainties, as represented in shaded area, were determined by dividing the data into $5$ equal sections.}
\label{fig7}
\end{figure}

In this section we applied our method to study the permeation of 
Domperidone, a dopamine receptor antagonist, across a POPC lipid membrane~\cite{2014-JCR-Eyer-Kramer,2017-JACS-Dickson-Duca,2018-JPCB-Badaoui-Rosta}, 
as illustrated in Fig.~\ref{fig7}(a). 
The simulation data were obtained from previous studies~\cite{2017-JACS-Dickson-Duca,2018-JPCB-Badaoui-Rosta} 
and further details of the simulation methods and parameters can be found there.
This system shows long time scale fluctuations inherent to the presence 
of the bilayer structure, which is representative of the transport of solute 
across heterogeneous media~\cite{2017-JACS-Dickson-Duca}.
As described in the original work of Dickson et al.~\cite{2017-JACS-Dickson-Duca}, 
the starting structures for the US simulations were obtained by placing the ligand 
at the center of a POPC bilayer surrounded by water molecules ($72$ POPC and $60$ 
waters per lipid). To generate the US windows, the drug was pulled out from the center 
of the system to outside the membrane, for a total of $40~\textrm{\AA}$. 
Configurations were saved every $1~\textrm{\AA}$, from the center $z=0~\textrm{\AA}$ 
to $z=40~\textrm{\AA}$ generating $40$ windows. 
Each US window was run for $20$ ns to allow equilibration, followed by additional $80$ ns 
of the production run using an US force constant of 
$2.5~\textrm{kcal}~\textrm{mol}^{-1}\textrm{\AA}^{-2}$.
\begin{figure*}[t]
\includegraphics[width=0.99 \textwidth, angle=-0]{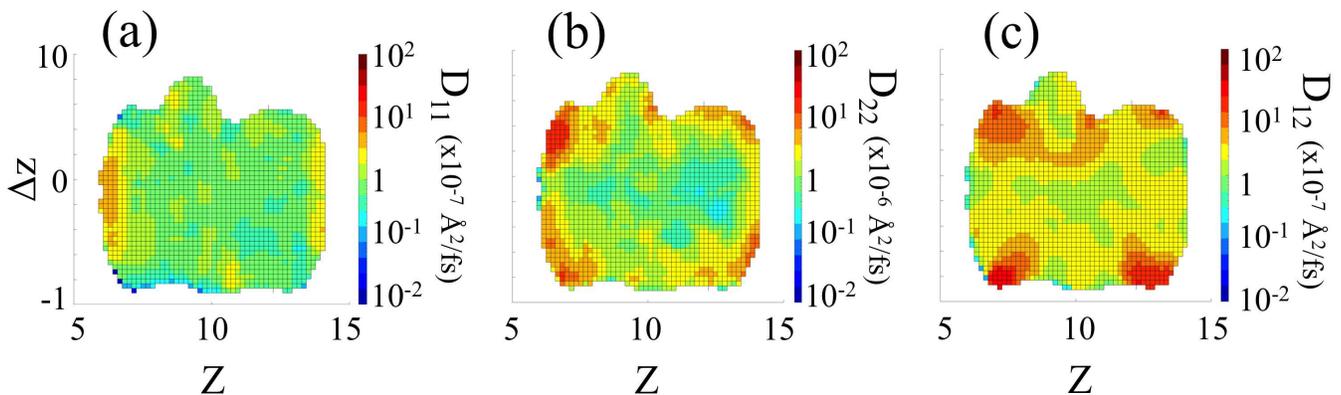}
 \caption{2D projection of the diagonal (a-b) and off-diagonal (c) elements 
 of the diffusion tensor, $D_{11}(Z,\Delta z)$, $D_{22}(Z,\Delta z)$, 
 and $D_{12}(Z,\Delta z)$, associated with the crossing of Domperidone across 
 the POPC membrane bilayer.
 We restrained the exploration of the free energy landscape to the area 
 centered around the minimal energy well ($Z \approx 10 \textrm{\AA}$ 
 in Figs.~\ref{fig7}(a) and (b)).
 }
\label{fig10}
\end{figure*}

The FEP and the diffusion profiles associated with the transport of Domperidone 
across the POPC bilayer, as shown in Fig.~\ref{fig7}(b) and (c), respectively,
were obtained using the DHAM.
As the drug reaches the water/bilayer interface, the value of the 
diffusion coefficient significantly increases before remaining approximately 
constant until the molecule reaches the adjoining area of the two lipid layers, 
which is representative of the hydrophobic nature of the compound.
These results can be interpreted within the inhomogeneous solubility-diffusion 
model commonly used to provide realistic description of membrane 
permeation~\cite{2016-BBA-Shinoda}. In this model, the permeability, Perm, 
is written as
\begin{equation}
\frac{1}{\textrm{Perm}}=\int_{-d/2}^{+d/2}\frac{1}{K(Z)D(Z)}=\int_{-d/2}^{+d/2}\frac{\exp\big(\Delta G(Z)/k_B T\big)}{D(Z)}~,
\label{eqPerm}
\end{equation}
where $K(Z)$, $D(Z)$, and $d$ are the position-dependent partition coefficient, the solute diffusion 
coefficient, and the membrane thickness respectively. $\Delta G(Z)$ is the free energy difference, 
which is related to the partition coefficient $K(Z)=\exp\big(-\Delta G(Z)/k_B T\big)$.
Considering the results shown in  Fig.~\ref{fig7}, we obtained the $\log$ Perm 
value associated with the transport of the Domperidone molecule across 
the POPC bilayer, $\log \textrm{Perm} = -2.71 \pm 0.15$,  which is in agreement 
with experimental results ($\log \textrm{Perm}^{(\textrm{exp})}=-2.6$)~\cite{2014-JCR-Eyer-Kramer,2017-JACS-Dickson-Duca,2018-JPCB-Badaoui-Rosta}.

\subsection{Membrane permeation in 2D}
We extended the 1D study above to the 2D analysis of the permeation of Domperidone 
across the POPC lipid membrane. To do so, we  considered the 
rotational movement of the drug during its passage across the membrane as 
additional degree of freedom.
We constructed a discretized 
two-dimensional grid to determine the MSMs along two reaction coordinates 
(see details in the SI). 
As our first reaction coordinate, we used the distance from the drug center 
of mass to bilayer center, $Z$, as already considered in 1D. 
For our second coordinate, we used the projection of the molecular 
orientation vector onto the axis orthogonal to the bilayer axis, $\Delta Z$, 
as a measure of the orientation of the drug with respect to the membrane, 
as described in Badaoui et al.~\cite{2018-JPCB-Badaoui-Rosta}.
We restrained the exploration 
of the free energy landscape to the area centered around the minimal 
energy well ($Z \approx 10 \textrm{\AA}$ in Figs.~\ref{fig7}(a) and (b)).
We measured the diagonal and off-diagonal KM coefficients 
defined in Eqs.~\ref{eqDriftDiffusion2D:1} and \ref{eqDriftDiffusion2D:2} 
(see details in the SI). Subsequently, the associated elements of the diffusion 
tensor were obtained with weighted polynomial regression.

In Fig.~\ref{fig10}(a), (b), and (c), we calculated the 2D projection of the diagonal 
and off-diagonal elements of the diffusion tensor, $D_{11}(Z,\Delta z)$, 
$D_{22}(Z,\Delta z)$, and $D_{12}(Z,\Delta z)$, respectively. 
We observed good agreement in the order of magnitude between the measure 
of $D_{11}(Z,\Delta z)$ (Fig.~\ref{fig10}(a)) and the one of $D^{(2)}(Z)$ 
(Fig.~\ref{fig7}(c)) in the minimal energy well. 
Additionally, we measured that the orientation of the drug with respect to the membrane 
diffuses an order of magnitude faster than the drug COM through the bilayer.
Finally, we estimated the off-diagonal element of the diffusion tensor, 
which is essential for understanding the importance of dynamic coupling 
between the coordinates. As shown in Fig.~\ref{fig10}(c), the order of magnitude 
of $D_{12}(Z,\Delta z)$ is comparable to the one measured for $D_{11}(Z,\Delta z)$ 
and $D_{22}(Z,\Delta z)$. 

\section{Conclusion}
In the present study, we presented a general, automatic method 
for estimating multi-dimensional position-dependent diffusion 
coefficients equally valid in biased or unbiased MD simulations. 
We combined Markov State Model analysis and Kramers-Moyal  expansion  
to link the underlying  stochastic process obtained within the 
non-linear Langevin framework to its probabilistic description. 
Our approach neither requires prior assumptions regarding the form 
of the free energy landscape nor additional numerical integration scheme.
We applied our numerical approach to one- and two-dimensional 
analytic potentials and data from explicit solvent molecular dynamics 
simulations, including the water-mediated conformations of the alanine 
dipeptide.\\ 

Importantly, we demonstrated the efficiency of our algorithm 
in studying the transport of solute across three-dimensional 
heterogeneous porous media, which is known to show long time 
scale fluctuations potentially breaking ergodicity in sampling~\cite{2016-JCIM-Lee-Gumbart}.
Specifically, our algorithm provided accurate assessment of the diffusion 
coefficient associated with the crossing of Domperidone 
across a lipid POPC membrane, which was not previously accessible with 
standard methods~\cite{2017-JACS-Dickson-Duca}.\\

The method would also allow the measure of the off-diagonal elements 
of the diffusion tensor, which is essential for assessing the importance 
of dynamic coupling between the reaction coordinates. As discussed by 
Ma and coworkers~\cite{2006-JCP-Ma-Dinner}, the presence of significant 
off-diagonal element in the diffusion tensor could indeed impact 
the determination of the reactive eigenvector at play in the Kramers' theory 
of reaction kinetics and then the measure of the rate associated with 
the kinetic evolution of the system.
This is particularly important in  toxicology  and pharmacology, where 
the prediction of kinetic quantities of solutes, including drugs and small peptides 
provides fundamental   understanding of numerous biochemical transport processes 
involved in the design of new drugs~\cite{2018-NRC-Amaro-Mulholland}.

\section{Acknowlegments}
We acknowledge use of the research computing facility at 
King’s College London, Rosalind (https://rosalind.kcl.ac.uk).
F.S. and E.R. thanks Gerhard Hummer for fruitful discussions and Magd Badaoui 
for his help in setting up the membrane bilayer system.
We acknowledge the support of the UK Engineering and Physical Sciences 
Research Council (EPSRC), under grant number EP/R013012/1, and ERC project 757850 BioNet.

\bibliography{acs}

\pagebreak
\widetext
\begin{center}
\textbf{\large Assessing Position-Dependent Diffusion from Biased Simulations 
and Markov State Model Analysis} \end{center}

\begin{center}\textbf{\large Supporting Information}
\end{center}
\setcounter{equation}{0}
\setcounter{figure}{0}
\setcounter{table}{0}
\setcounter{page}{1}
\makeatletter
\renewcommand{\theequation}{S\arabic{equation}}
\renewcommand{\thefigure}{S\arabic{figure}}

\section*{1D Langevin Dynamics}
\textbf{Analytical Potentials.} In the model systems evolving under 
the Brownian overdamped and the full inertial limit of the Langevin equation 
in 1D, the deterministic force, $F(x)=-\nabla V(x)$, in Eqs.~14 and 18 in the 
main text derives from the potential energy 
$V(x) = V_{\textrm{ref}}(x) + V_{\textrm{bias}}(x)$, where $V_{\textrm{ref}} = \sum_{n=0}^{6} \alpha(n) x^n$, 
is defined as a polynomial of degree $6$, with parameters $\alpha(n)$ given below for 
the low ($V_{\textrm{low}}$) and high ($V_{\textrm{high}}$) energy barrier potentials:
\begin{table}[h]
    \centering
    \begin{tabular}{c|c|c|c|c|c|c|c}
    \hline
     $V_{\textrm{ref}}$ &  $\alpha(0)$ & $\alpha(1)$ & $\alpha(2)$ & $\alpha(3)$ & $\alpha(4)$ & $\alpha(5)$ & $\alpha(6)$\\
     \hline\hline
     $V_{\textrm{low}}$ &  $22.7498$ &  $-301.374$ &  $1.3865\times 10^3$ &  $-2.9683\times 10^3$ &  $3.2173\times 10^3$ &  $-1.7111\times 10^3$ &  $354.3680$ \\
     \hline
     $V_{\textrm{high}}$ &  $47.2598$ & $-722.6061$ & $3.8067\times 10^3$ & $-8.7222\times 10^3$ & $9.7780\times 10^3$ & $-5.2710\times 10^3$ & $1.0919\times 10^3$ \\
     \hline
    \end{tabular}
\end{table}
\newline
The biased energy potential, $V_{\textrm{bias}} = \frac{1}{2} K \big(x-x^{(k)}\big)^2$  
is defined as an harmonic potential with $K$ the biasing spring constant and $x^{(k)}$ 
the center of the harmonic bias in simulation $k$, uniformly distributed 
along the reaction coordinate (RC). \\

\textbf{Analytical Diffusion coefficients.} We considered either a quadratic or step-like position-dependent diffusion 
profile, $D(x) = k_BT/\gamma(x)$ defined by the generalization of Einstein's relation, 
with $k_B$ the Boltzmann constant, $T=300$ K the temperature of the system, 
and $\gamma(x)$ the position-dependent friction defined as a parabolic (P) 
or a $Z$-shaped membership (Z) function,
\begin{eqnarray}
    \gamma^{P}(x) &=& \gamma_0^P \Big( 1-\frac{1}{3}(x-x_P)^2 \Big) \,, \\
    \gamma^{Z}(x) &=& \gamma_0^Z \Big( 2+\textrm{zmf}(x,a,b) \Big) \,,
\end{eqnarray}
with $\gamma_0^P = 3000$, $x_P=0.8$, $\gamma_0^Z=1700$, $a=0.5$, $b=1.1$.
The function $\textrm{zmf}(x,a,b)$ represents the sigmoidal membership function defined as
$$
\textrm{zmf}(x,a,b) = \left\{
    \begin{array}{ll}
        1 \,, & x \leq a \\
        1-2\Big(\frac{x-a}{b-a}\Big)^2 \,, & a\leq x \leq \frac{a+b}{2} \\
        2\Big(\frac{x-b}{b-a}\Big)^2 \,, &  \frac{a+b}{2}\leq x \leq b \\
        0 \,, & x\geq b
    \end{array}
\right.
$$
\newpage

\begin{figure}[h]
\includegraphics[width=0.4 \textwidth, angle=-0]{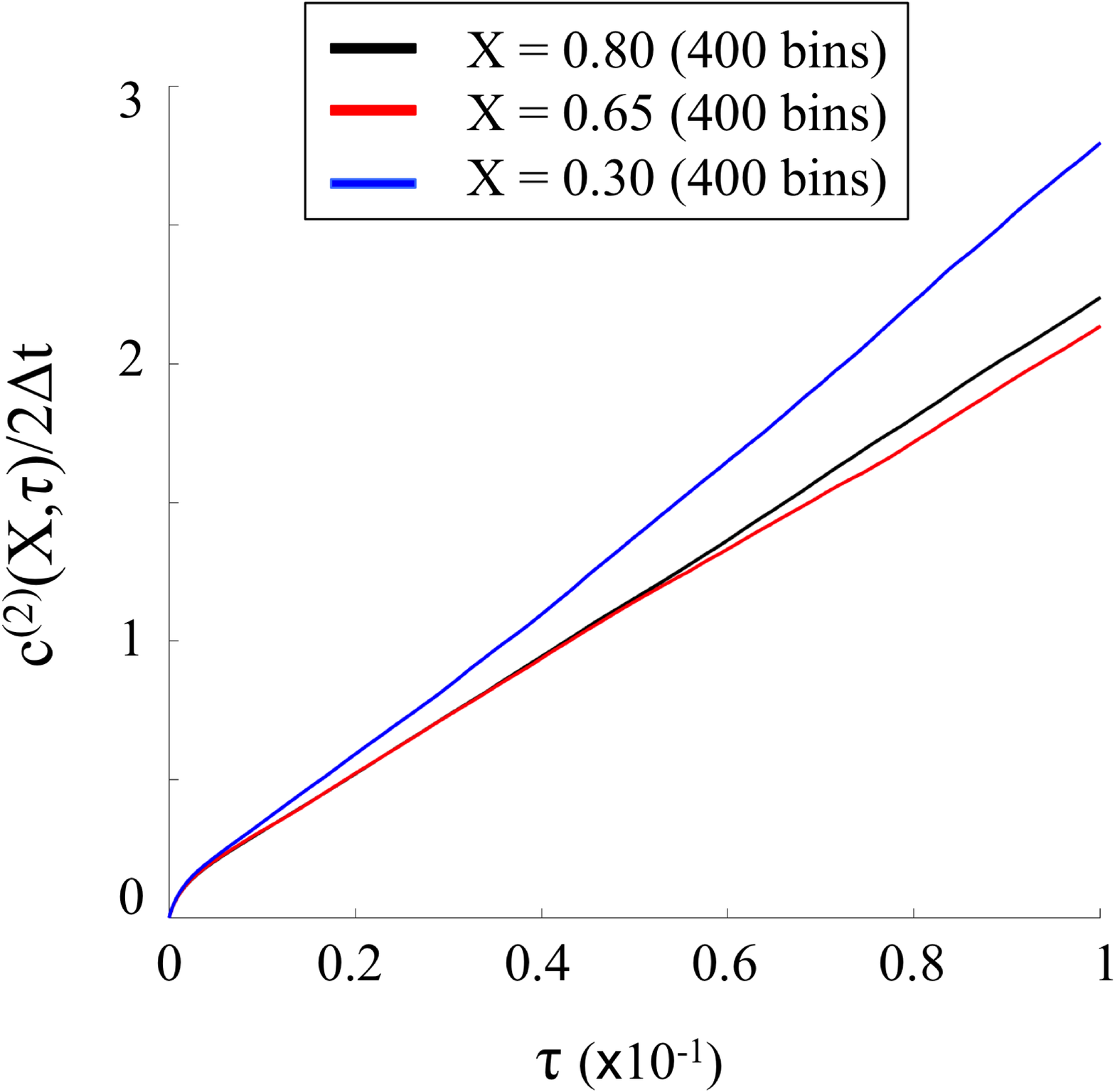}
 \caption{Evolution of $c^{(2)}(X,\tau)$ in the  biased full inertia Langevin 
 dynamics and $V_{\textrm{high}}$ for three different position along the RC. 
 The discretization of the system considered here was $N_{\textrm{bin}}=400$.}
\label{figS1}
\end{figure}

\begin{figure}[h]
\includegraphics[width=0.8 \textwidth, angle=-0]{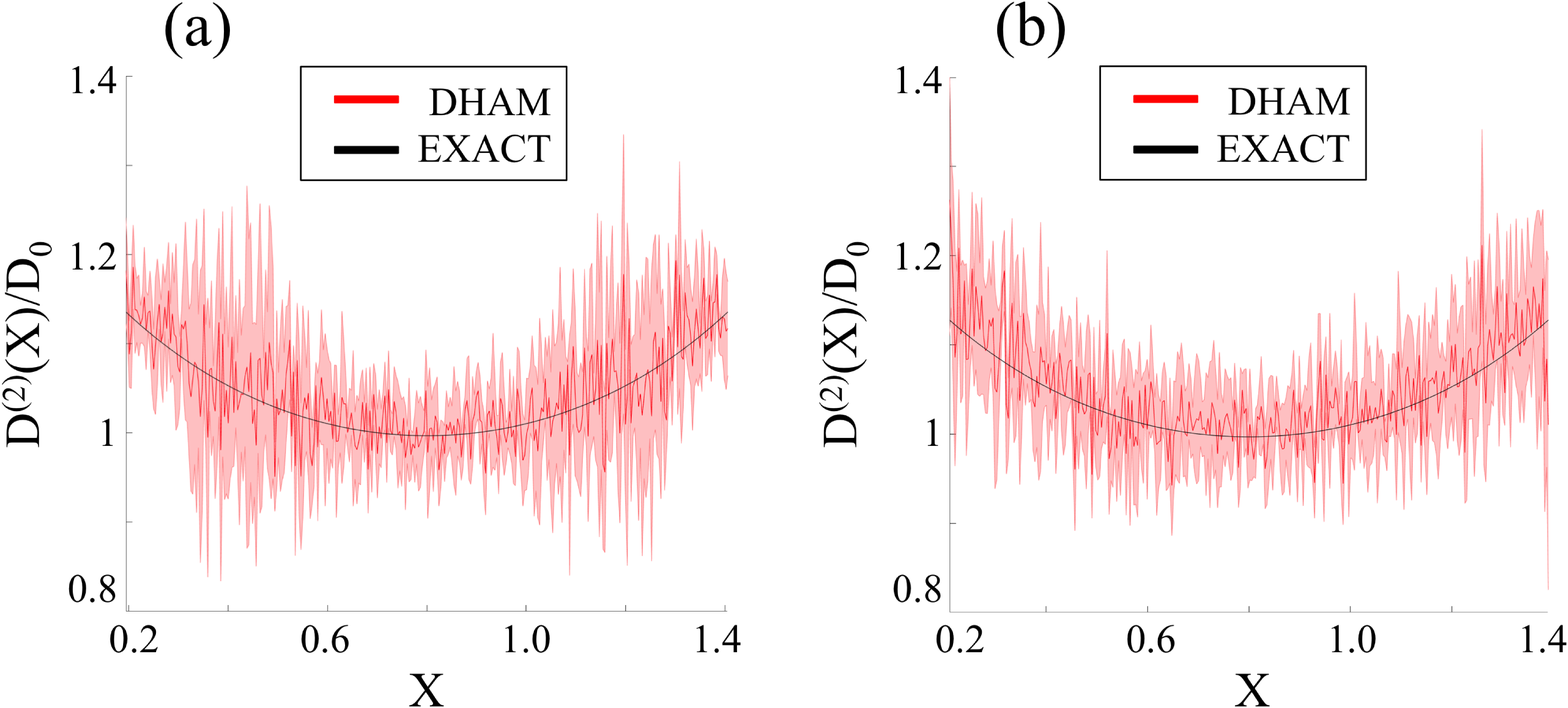}
 \caption{Diffusion profiles obtained with biased full inertia Langevin simulations in 
 $V_{\textrm{high}}$ and quadratic diffusion when the biasing spring constant 
 $K=500$ kcal/mol (a) and $K=2000$ kcal/mol (b). The Diffusion coefficient $D^{(2)}(X)$ 
 is rescaled to $D_0 = k_BT/\gamma_0^P$ for clarity. The discretization of the system 
 considered here was $N_{\textrm{bin}}=400$. The profiles obtained within the DHAM 
 framework (red) agree within standard error with the exact profiles obtained from 
 the generalization of Einstein's relation (black). Uncertainties, as represented 
 in shaded area, were estimated from 10 independent runs, determining the profiles 
 independently, and calculating the standard error.}
\label{figS2}
\end{figure}

\textbf{Biased 1D Langevin dynamics in $V_{\textrm{high}}$.} We considered 
the biased evolution of the system in the high $(V_{\textrm{high}})$ barrier 
potential defined above, within the umbrella sampling framework. We studied the effect 
of increasing the biasing spring constant, $K$, from $500$ kcal/mol to $2000$ kcal/mol, on 
the reconstruction of the diffusion profile. 
We first evaluated the second conditional moments $c^{(2)}$, defined 
in Eq.~$4$ in the main text, as shown in Fig.~\ref{figS1}.
%
The diffusion coefficient was subsequently assessed from the slope of a weighted 
polynomial regression. As shown in Fig.~\ref{figS2}, we observed excellent agreement 
between the theoretical profiles and the numerical results. For the lowest biasing 
spring constant, $K=500$ kcal/mol, we oberved higher variability 
around the transition states $X \approx 0.5$ and $\approx 1.2$, as shown 
in Fig.~\ref{figS2} (a).
%
Increasing the biasing spring constant, $K$, from 
$500$ kcal/mol to $2000$ kcal/mol, yields better sampling around the transition states, 
with lower variability on the reconstruction of the diffusion profile around 
the transition states, as shown in Fig.~\ref{figS2} (b).

\section*{2D Langevin Dynamics}
\begin{figure}[h]
\includegraphics[width=0.6 \textwidth, angle=-0]{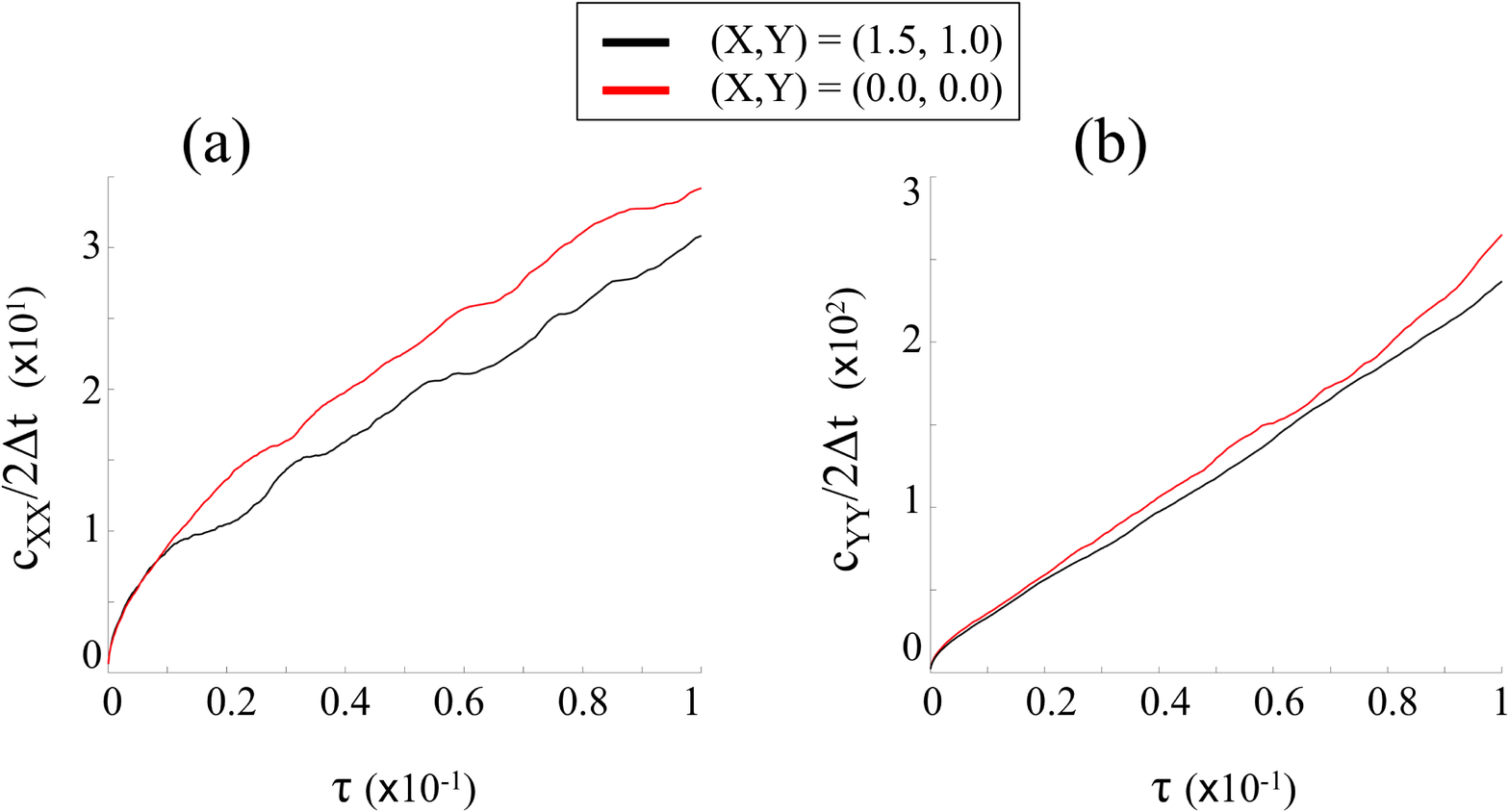}
 \caption{Evolution of the diagonal conditional moments $c_{XX}^{(2)}$ (a) 
 and $c_{YY}^{(2)}$ (b) in the  biased Brownian overdamped Langevin dynamics 
 in 2D, for two different positions in the potential energy surface 
 $(X,Y)=(1.5, 1.0)$ and $(X,Y)=(0, 0)$ and $N_{\textrm{bin}}=90 \times 90$.}
\label{figS3}
\end{figure}

Analogously to the 1D case, we constructed a finely discretized 2D grid 
to determine the MSMs along the two RC $X$ and $Y$ using the 2D-DHAM method.
We considered a 2D binning of $90 \times 90$, sufficiently small to measure 
accurately the diagonal elements of the diffusion tensor, $D_{11}$ and $D_{22}$, 
shown in Fig.~6 in the main text. 
To do so, we first evaluated the diagonal conditional moments, $c_{XX}$ and $c_{YY}$, 
defined in Eq.~$13$ in the main text, as shown in Figs.~\ref{figS3} (a) and (b), 
respectively.
The diffusion coefficient was subsequently assessed from the slope of a weighted 
polynomial regression.

\section*{Water-mediated conformations of Alanine Dipeptide in 1D}
\begin{figure}[b]
\includegraphics[width=0.57 \textwidth, angle=-0]{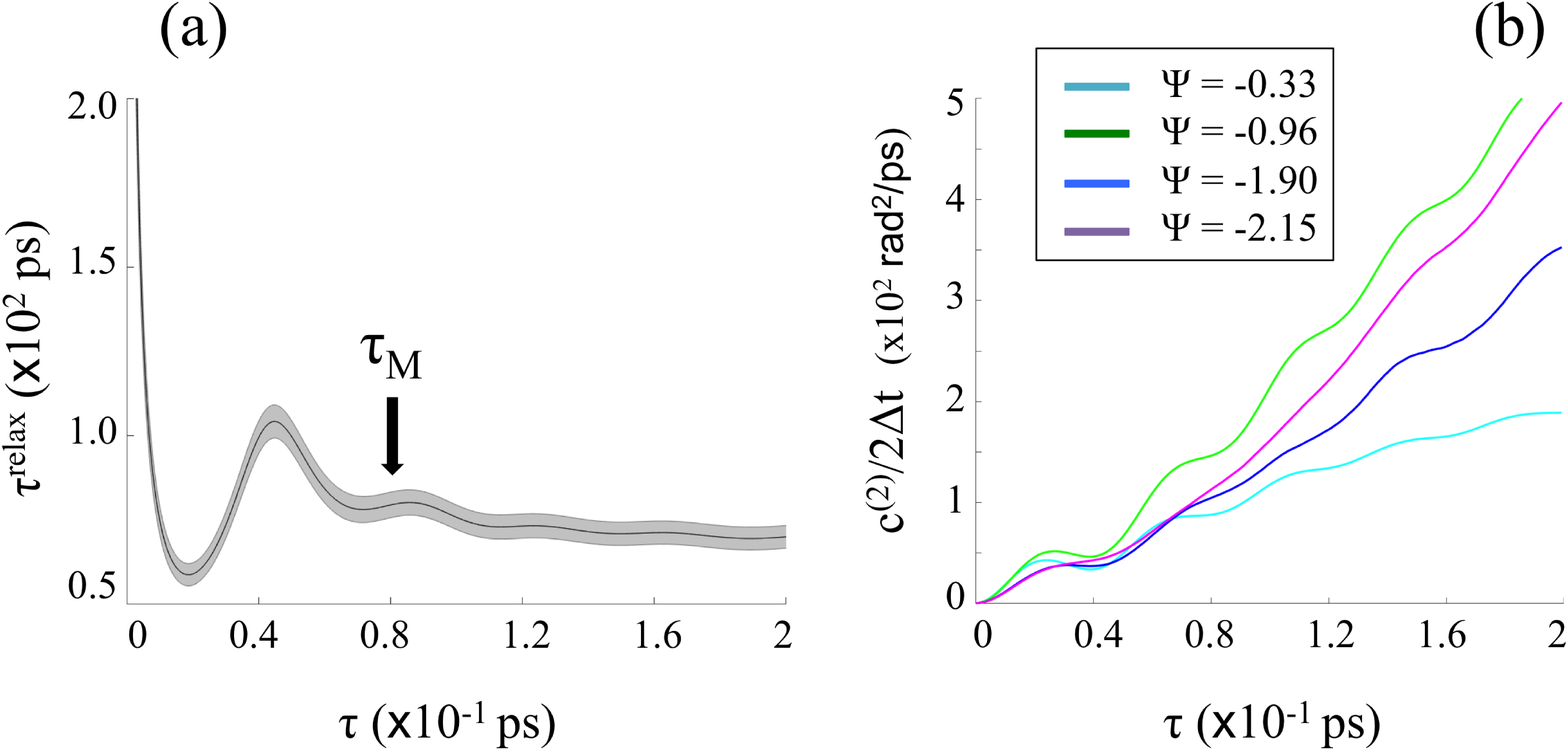}
 \caption{Evolution of (a) the slowest relaxation time, $\tau^{\textrm{relax}}$, 
 and (b) the conditional moment $c^{(2)}$ for different values of the dihedral angle 
 $\Psi$ measured for $N_{\textrm{bin}}=1000$. Uncertainties, as represented 
 in shaded area, were estimated from 3 independent runs, determining the profiles 
 independently, and calculating the standard error.}
\label{figS4}
\end{figure}
To obtain the diffusion profile associated with the conformational transition 
between the metastable states $\alpha$ and $\beta$ of 
ALA2 shown in Fig.~7 in the main text, we assessed the Markov timescale 
$\tau_M$ associated with the MD trajectory for 
a discretization $N_{\textrm{bin}}=1000$ (Fig.~\ref{figS4} (a)). 
We compared our results with those obtained with a different discretization 
$N_{\textrm{bin}}=500$, which did not show significant changes (data not shown).
We then evaluated the second conditional moments, $c^{(2)}$, over the range 
of lagtime $\tau \geq \tau_M$, as shown in Fig.~\ref{figS4} (b). The diffusion coefficient 
was assessed from the slope of a weighted polynomial regression.
We calculated the fourth-order coefficient and check the ratio 
$D^{(4)}(X)/\big(D^{(2)}(X)\big)^2 < 10^{-2}$, indicating the necessary condition 
of the Pawula theorem holds.

\section*{Membrane permeation}
\begin{figure}[h]
\includegraphics[width=0.7 \textwidth, angle=-0]{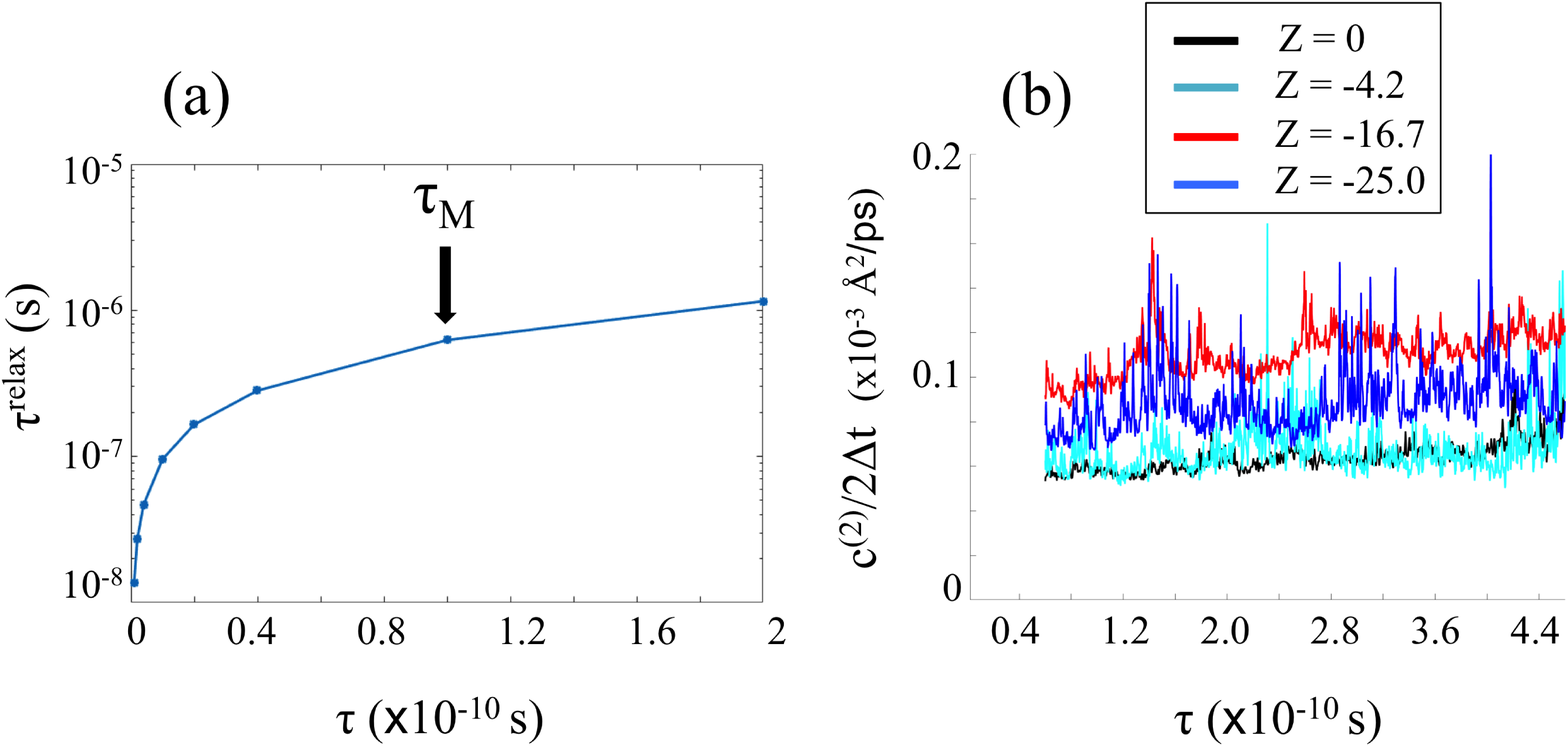}
 \caption{Evolution of (a) the slowest relaxation time, $\tau^{\textrm{relax}}$, 
 and (b) the conditional moment $c^{(2)}$ for different values of the position $Z$ 
 of the ligand through the POPC membrane measured for $N_{\textrm{bin}}=1000$.}
\label{figS5}
\end{figure}

\begin{figure}[h]
\includegraphics[width=0.45 \textwidth, angle=-0]{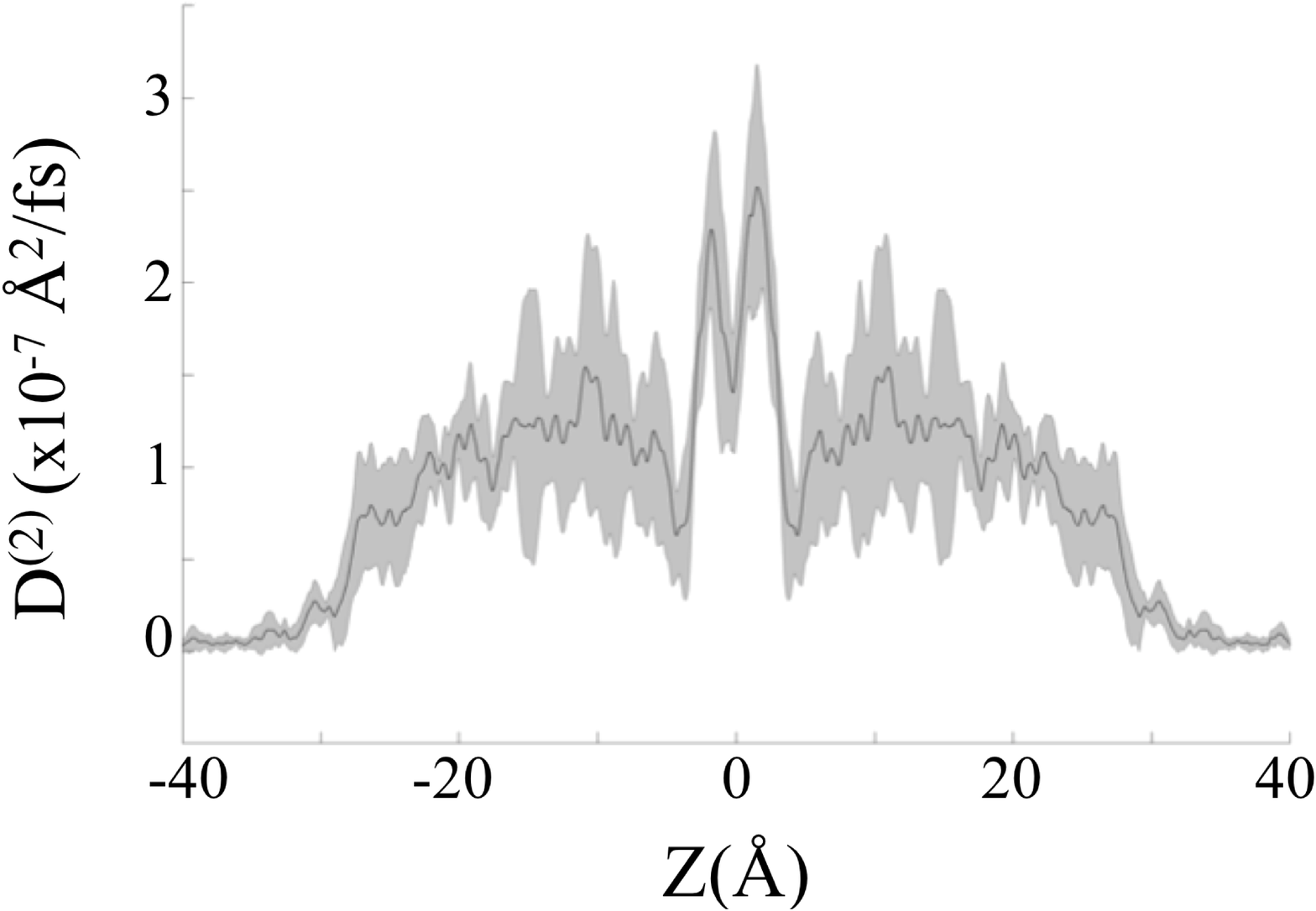}
 \caption{Diffusion profile associated with the crossing of Domperidone through 
 the (POPC) bilayer and reconstructed along the distance from the center 
 of the bilayer, $Z$, measured for $N_{\textrm{bin}}=1000$. 
 Uncertainties, as represented in shaded area, were determined by dividing 
 the data into two equal sections, determining the profiles independently, 
 and calculating the standard error.}
\label{figS6}
\end{figure}

To obtain the 1D diffusion profile associated with the permeation of 
Domperidone, a dopamine receptor antagonist, across the lipid (POPC) membrane 
shown in Fig.~8 in the main text, we first assessed the Markov timescale $\tau_M$ associated 
with the MD trajectory for a discretization $N_{\textrm{bin}}=1000$ (Fig.~\ref{figS5} (a)).
The relaxation time can be seen to level off in the region of lagtimes greater 
than 100 ps, where it is sufficiently large for the relaxation time to be insensitive 
to the precise choice of the lagtime, as shown in Fig.~\ref{figS5} (a).
We then evaluated the second conditional moments $c^{(2)}$ defined 
in Eq.~$4$ in the main text over the range of lagtime $\tau \geq \tau_M$, 
as shown in Fig.~\ref{figS5} (b). The diffusion profile, as shown in Fig.~\ref{figS6}, 
was subsequently assessed from the slope of a weighted polynomial regression. 
We calculated the fourth-order coefficient and check the ratio 
$D^{(4)}(X)/\big(D^{(2)}(X)\big)^2 < 10^{-2}$, 
indicating the necessary condition of the Pawula theorem holds (data not shown).

\begin{figure}[h]
\includegraphics[width=0.4 \textwidth, angle=-0]{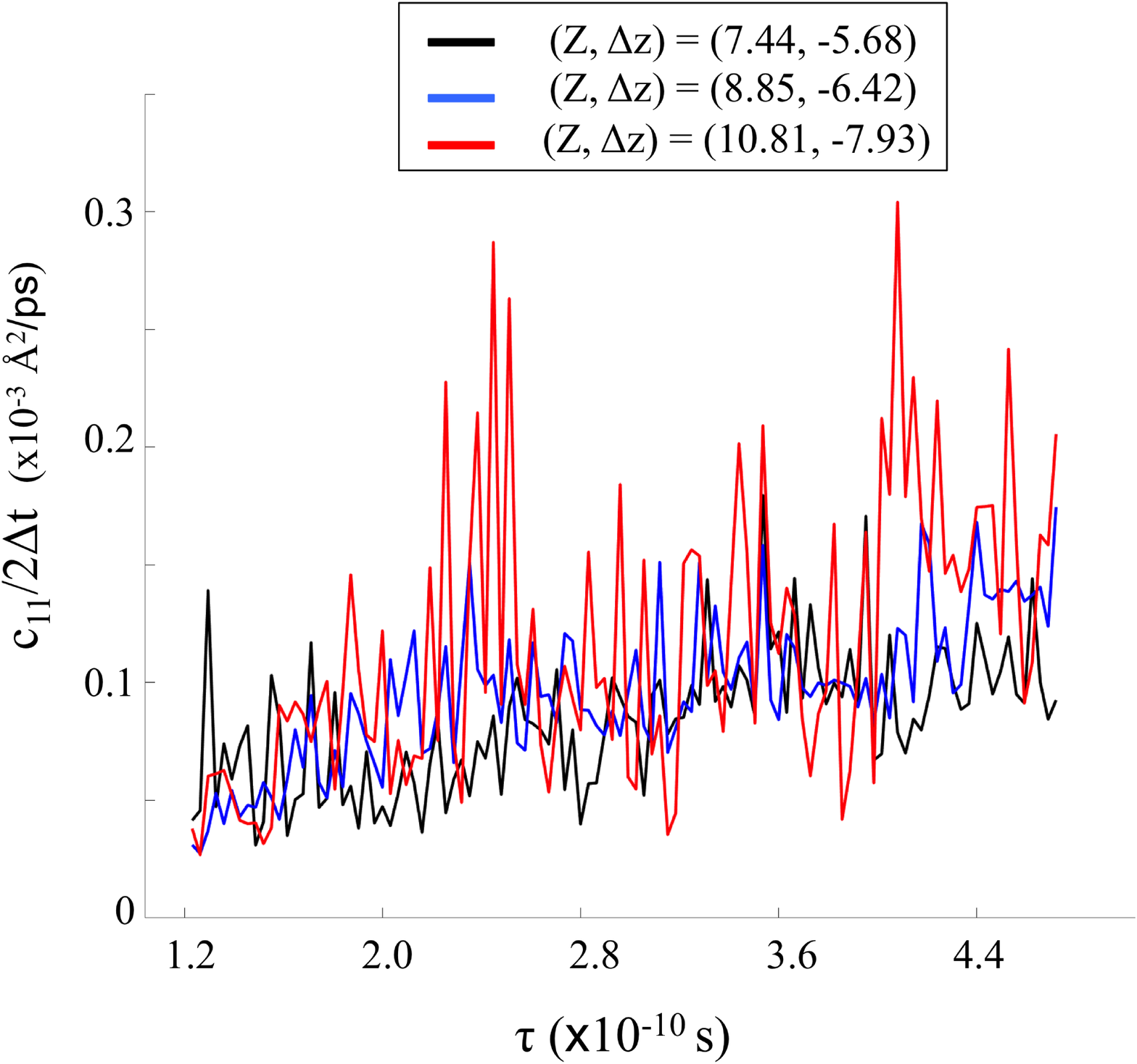}
 \caption{Evolution of the diagonal conditional moments $c_{11}^{(2)}$ obtained 
 in the analysis of the membrane permeation in 2D for three different positions 
 in the potential energy surface $(Z,\Delta z)$ and $N_{\textrm{bin}}=50 \times 50$.}
\label{figS7}
\end{figure}

Analogously to the 1D case, we constructed a finely discretized 2D grid to 
determine the MSMs along two reaction coordinates, $Z$ and $\Delta z$, using 
the 2D-DHAM method. 
We considered a 2D binning of $50 \times 50$, sufficiently small to 
evaluate the second conditional moments (see Fig.~\ref{figS7}), and measured  
the diagonal and off-diagonal elements of the diffusion tensor in the localized 
area defined in Fig.~9 in the main text.
We calculated the fourth-order coefficient and check the ratio 
$D_{ii}^{(4)}(X)/\big(D_{ii}^{(2)}(X)\big)^2 < 2\times 10^{-2}$, 
indicating the necessary condition of the Pawula theorem holds.

\end{document}